\documentclass[12pt]{article}
\RequirePackage[margin=1in,top=1in,bottom=1in]{geometry} 

\usepackage{setspace}

\usepackage{verbatim}
\usepackage[font=small]{caption}

\usepackage{multirow}
\usepackage{amsmath,amsthm,amssymb,url}
 
\usepackage{graphicx}

\usepackage{color}
\usepackage{tabu}
\usepackage{natbib}
\usepackage{tikz}
\usepackage{cprotect}
\usetikzlibrary{positioning}

\tikzset
{
node/.style={draw, rectangle, align=center, minimum size=1cm}
}
\newdimen\nodeDist
\newcommand{\Real}{\mathbb{R}}

\title{{Regression Trees for Cumulative Incidence Functions}}
\author{Youngjoo Cho, Annette M. Molinaro, \\
Chen Hu, and Robert L. Strawderman$^*$ \\[2ex]
$^*$ robert\_strawderman@urmc.rochester.edu \\
Department of Biostatistics \& Computational Biology  \\
601 Elmwood Avenue, Box 630 \\
University of Rochester \\
Rochester NY 14642}

\date{}

\begin{document}
\singlespacing
\maketitle


\begin{abstract}
The use of cumulative incidence functions for characterizing the risk of one type of event in the presence of others has become increasingly popular over the past decade. The problems of modeling, estimation and inference have been treated using parametric, nonparametric and semi-parametric methods. Efforts to develop suitable extensions of machine learning methods, such as regression trees and related ensemble methods, have begun only recently. In this paper, we develop a novel approach to building regression trees for estimating cumulative incidence curves in a competing risks setting. The proposed methods employ augmented estimators of the Brier score risk as the primary basis for building and pruning trees. The proposed methods are easily implemented using the 
R statistical software package. Simulation studies demonstrate the utility of our approach in the competing risks setting. Data from the Radiation Therapy Oncology Group (trial 9410) is used to illustrate these new methods.\\

\noindent \textit{Keywords: Brier score; CART; Cause-specific hazard; Fine and Gray model; Sub-distribution function}
\end{abstract}
 
\newpage

\doublespacing
\setstretch{1.8} 

\section{Introduction}

A subject being followed over time may experience several types of events related, for example, to disease morbidity and mortality. For example, in a Phase III trial of concomitant versus sequential chemotherapy and thoracic radiotherapy for patients with inoperable non-small cell lung cancer (NSCLC) conducted by the Radiation Therapy Oncology Group (RTOG), patients were followed up to 5 years, the occurrence of either disease progression or death being of particular interest. Such ``competing risks'' data are commonly encountered in cancer and other biomedical follow-up studies, in addition to the potential complication of right-censoring on the event time(s) of interest. 

Two quantities are often used when analyzing competing risks data: the cause-specific hazard function (CSH) and the cumulative incidence function (CIF). For a given event, the former describes the instantaneous risk of this event at time $t$, given that no events have yet occurred; the latter describes the probability of occurrence, or absolute risk, of that event across time and can be derived directly from
the subdistribution hazard function \citep{fine1999proportional}.
The literature on competing risks is growing rapidly, particularly for hazard-based regression modeling; see \citet{dignam} for a contemporary review, where parametric and semi-parametric approaches to modeling both the CSH and CIF are considered. The literature on tree-based methods for competing risks remains much less developed.  \citet{leblanc1992relative} extended the classification and regression tree (CART) procedure of \cite{breiman1984classification} to the case of right-censored survival data by replacing the squared error loss function ordinarily used for a continuous outcome with an approximation to the likelihood function derived under a certain proportional hazards assumption.
Motivated by methods for handling monotone missing data, \citet{molinaro2004tree} later proposed using an inverse 
probability weighted (IPCW) squared error loss function to directly generalize the original CART procedure for
right-censored survival data, obtaining a procedure that reduces to the standard form of CART for a continuous outcome when censoring is
absent.   \cite{steingrimsson2016doubly} recently proposed a ``doubly robust'' extension
motivated by semiparametric efficiency theory for missing data 
\citep{robins1994estimation,tsiatis2007semiparametric}, 
and demonstrated significantly improved performance in comparison to the IPCW-based tree procedure of \citet{molinaro2004tree}. 
However, such methods cannot be used when there is more than one type of event (i.e., in addition
to censoring). 
For two or more competing risks,  \cite{callaghan2008} proposed two methods of building classification trees for the CIF: by maximizing 
between-node heterogeneity via the two-sample log-rank test of \cite{gray1988class}; 
 and, by maximizing within-node homogeneity using sums of event-specific martingale residuals. 
To our knowledge, no software package is available that implements these methods,
and more importantly, there is also no software specifically targeted
to the problem of fitting regression trees to competing risk outcomes.
In contrast, ensemble methods for estimating the CIF as a function of covariates do exist; see, for example,
\cite{ishwaran2014random} and \cite{mogensen2013random}. \cite{ishwaran2014random} implement
their methods in the {\sf randomForestSRC} package \citep{rfsrc},
where the unpruned trees that make up the bootstrap ensemble are typically built  using
splitting criteria designed for group comparisons with a single competing risk outcome, 
such as the generalized logrank test.


This paper proposes a novel CART-based approach to building regression trees under competing risks.
In Section \ref{data}, we introduce the relevant data structures and describe a direct approach to building regression trees 
for estimating the CIF for a specified cause  when there 
is no loss to follow-up. Section \ref{Stein-gen} develops the necessary extensions of these loss functions for right-censored 
outcomes, 
along with extensions for estimating the CIF at several time points. These methods focus on directly estimating the CIF,
avoiding estimation of the cause-specific or subdistribution hazard function. In Section \ref{sims}, simulation studies are used
to investigate performance of these new methods and Section \ref{RTOG9410} summarizes analysis results for 
the RTOG 9410 Phase III lung cancer trial mentioned at the beginning of this section. The paper concludes with comments
on future work in Section \ref{discuss}.


\section{CIF Regression Trees with Uncensored Data}
\label{data}
This section introduces the relevant competing risks data structure and reviews a CART-based approach to building regression trees for estimating the CIF  in the case where there is no other loss to follow-up (i.e., no random censoring). This necessary background will allow us to develop an analogous approach when there is loss to follow-up.

\subsection{Full Data Structure}
\label{S:data}
Let $T^{(m)}$ be the time to failure for failure type $m = 1,\ldots, K, K \geq 2$ and let $W$ be a vector of $p$ covariates, where $W \in {\cal S} \subset \Real^p$. Let $T = \min(T^{(1)},\ldots, T^{(K)})$ be the minimum of all latent failure times; it is assumed $(T^{(1)},\ldots, T^{(K)})$ has a bounded multivariate density function and hence that $T$ is continuous. Then, in the absence of other loss to follow-up, $F= (T,W,M)$ is assumed to be the fully observed (or full) data for a subject, where $M$ is the event type corresponding to $T.$ This assumption implies that  $(T^{(M)}, M, W)$ is observed and, in addition, that $T^{(m)} > T^{(M)}$ for every $m \neq M;$ however, $T^{(m)}$ itself is not assumed to be observed for $m\neq M$. Let $\mathcal{F} = (F_1,\ldots,F_n)$ be the full data observed on $n$ independent subjects,  where $F_i=(T_i,W_i,M_i), i=1,\ldots,n$ 
are assumed to be identically distributed (i.i.d.). The dependence
structure of $(T^{(1)},\ldots, T^{(K)})$ is left unspecifed.
%


\subsection{CIF Estimation Under a Brier Loss Function}
\label{CIF-full}
Let $\Psi_0 = \{ \psi_{0m}(t;w) = P(T \leq t, M = m | W=w), t > 0, m=1,\ldots,K\}.$ The set of CIFs $\Psi_0$ can be estimated from the data ${\cal F}$ using any of several suitable parametric or semiparametric methods without further assumptions (e.g., independence of $T^{(1)},\ldots, T^{(K)}$).  This section introduces a loss-based method for estimating $\psi_{0m}(t;w)$ for a fixed cause $m$ and time point $t >0$ 
in the case where $\psi_{0m}(t;w)$ is piecewise constant as a function of $W.$ This is a key step in our proposed
approach to building a regression tree to estimate $\psi_{0m}(t;w).$ 

Let $\mathcal{N}_1,\ldots,\mathcal{N}_L$ form a partition of ${\cal S}.$ 
Define $\beta_{0lm}(t) = P(T \leq t, M=m | W \in \mathcal{N}_l)$
and suppose
$\psi_{0m}(t;w) = \sum_{l=1}^{L}\beta_{0lm}(t)I\{W \in \mathcal{N}_l\};$ that is, subjects falling
into partition $\mathcal{N}_l$ all share the same CIF $\beta_{0lm}(t).$
Define $Z_m(t) = I(T \leq t, M=m)$
and let  $\psi_{m}(t;w) = \sum_{l=1}^{L}\beta_{lm}(t)I\{W \in \mathcal{N}_l\}$
be a model for $\psi_{0m}(t;w).$ Then, 
fixing both $t > 0$ and $m,$ 
the Brier loss \citep[cf.,][]{brier1950verification} 
$
L_{m,t}^{full}(F,\psi_m) = \{Z_m(t) - \psi_m(t;w)\}^2
 = \sum_{l=1}^{L} I\{W \in \mathcal{N}_l\} \{Z_{m}(t) - \beta_{lm}(t)\}^2
$
is an unbiased estimator of the risk $\Re(t,\psi_m) = E[\sum_{l=1}^{L} I\{W \in \mathcal{N}_l\}\{Z_{m}(t) - \beta_{lm}(t)\}^2],$
or equivalently, $\Re(t,\psi_m) = \sum_{l=1}^{L} P\{W \in \mathcal{N}_l\}\{ \beta_{0lm}(t) - \beta_{lm}(t)\}^2.$
Assuming that ${\cal F}$ is observed, 
\begin{gather}
\label{eq:0}
L_{m,t}^{emp}({\cal F},\psi_m) = \frac{1}{n} \sum_{i=1}^n L_{m,t}^{full}(F_i,\psi_m)
= 
\frac{1}{n} \sum_{i=1}^n \sum_{l=1}^{L} I\{W_i \in \mathcal{N}_l\} \{Z_{im}(t) - \beta_{lm}(t)\}^2
\end{gather}
is also an unbiased estimator of $\Re(t,\psi_m).$ When considered as a function of the set of scalar parameters  
$\beta_{lm}(t), l =1,\ldots,L$ (i.e., for fixed $m$ and $t$), the empirical Brier loss 
\eqref{eq:0} is minimized when
$\psi_m(t;w) = \hat{\psi}_m(t;w) = \sum_{l=1}^{L} I\{W_i \in \mathcal{N}_l\} \hat{\beta}_{lm}(t),$ where
\begin{gather}
\label{beta-full}
 \hat{\beta}_{lm}(t) = \frac{ \sum_{i=1}^n  I\{W_i \in \mathcal{N}_l\} Z_{im}(t)}{\sum_{i=1}^n  I\{W_i \in \mathcal{N}_l\} }
\end{gather}
is a nonparametric estimate for $\beta_{0lm}(t), m=1,\ldots,K.$

\subsection{Estimating a CIF Regression Tree Using CART}
\label{implement0}

The CART algorithm of \citet{breiman1984classification} estimates a regression tree as follows:
\begin{enumerate}
\item Using recursive binary partitioning, grow a maximal tree by selecting a (covariate, cutpoint) combination at every
stage that minimizes an appropriate loss function;
\item Using cross-validation, select the best tree from the sequence of candidate trees generated by Step 1
via cost complexity pruning (i.e., using penalized loss).
\end{enumerate}
In its most commonly used form, CART estimates the conditional mean response as a piecewise constant function
on the covariate space, making all decisions on the basis of minimizing squared error loss. The resulting tree-structured regression
function estimates the predicted response in each terminal node using the sample mean of the observations falling into that node. 
This process generalizes in a straightforward way to more general loss functions.

In the absence of censoring and under a piecewise constant model 
\begin{gather}
	\label{psi mod}			
	\psi_m(t;w) = \sum_{l=1}^{L}\beta_{lm}(t)I\{W \in \mathcal{N}_l\}
\end{gather}  
for $\psi_{0m}(t;w),$ Section \ref{CIF-full} shows that a nonparametric 
estimate for $\psi_{0m}(t;w)$ for a given cause $m$ at a fixed $t > 0$ can be obtained by minimizing the loss \eqref{eq:0},
a problem equivalent to estimating the conditional mean response from the modified dataset
${\cal F}_{red,t} = \{ (Z_{im}(t), W'_i)',i=1,\ldots n \}$ by minimizing the squared error loss \eqref{eq:0}. Thus,
any implementation of CART for squared error loss (e.g.,  {\tt rpart}) applied to ${\cal F}_{red,t}$ will produce a regression tree estimate of $\psi_{0m}(t;w).$ In particular, CART estimates $L$ and the associated terminal nodes  $\{\mathcal{N}_1,\ldots,\mathcal{N}_L\}$ from the data ${\cal F}_{red,t},$ and within each terminal node, estimates $\psi_{0m}(t|w)$ by \eqref{beta-full}. While statistically inefficient, this process can be repeated for each $m=1,\ldots,K$
and any $t >0$ to generate nonparametric estimates of the CIF at a given set of time points for every cause.


\section{CIF Regression Trees with Censored Data}
\label{Stein-gen}

With ``full data'' the squared error loss plays a critical role in both steps of the CART algorithm described at the beginning
of Section \ref{implement0} when estimating a conditional mean \citep[e.g.,][Chapter 3.3]{breiman1984classification}. 
In follow-up studies involving competing risks outcomes, the full data ${\cal F}$ might not be observed due to loss to follow-up. In this case, estimating
$\psi_{0m}(t;w)$ for a specified $m$ under the squared error loss \eqref{eq:0} is not possible and one cannot run the CART procedure of
Section \ref{implement0} as described.

An attractive way to overcome this difficulty is to use to construct an observed data loss function that has the same risk as the 
desired full data loss function 
\citep[c.f.,][]{molinaro2004tree,lostritto2012partitioning,steingrimsson2016doubly}, that is,
the empirical Brier loss \eqref{eq:0}.
This section extends the developments of Section \ref{data} to the case of right-censored competing risks by deriving several new 
observed data loss functions that share the same risk as \eqref{eq:0}.
By substituting any of these new loss functions in for \eqref{eq:0} everywhere throughout the CART algorithm
of Section \ref{implement0}, one obtains a new regression tree method for estimating $\psi_{0m}(t;w).$ 
Similarly to \citet{steingrimsson2016doubly}, direct implementation  is possible using {\tt rpart}, 
which provides users with the ability to customize loss functions and decision making procedures. 
An appealing feature of the algorithms induced by these new loss functions is that each also reduces to the 
algorithm described in Section \ref{implement0} when censoring is absent.

\subsection{Observed Data Structure}

Let $C$ be a continuous right-censoring random variable that, given $W$, is statistically independent of  $(T,M).$ 
For a given subject, assume that instead of $F$ we only observe $O = \{\tilde{T},\Delta, M \Delta, W  \},$ where 
$\tilde{T} = \min(T,C)$ and $\Delta = I(T \leq C)$ is the (any) event indicator. The observed data on $n$ i.i.d.\ subjects is  
$\mathcal{O} = (O_1,\ldots,O_n).$ Similarly to the case where $K=1,$ random censoring on $T$ permits estimation of
the CIF from  the data ${\cal O}$. 

\subsection{Estimating $\Re(t,\psi_m)$ via IPCW Loss }
\label{CIF-obs}

Fix $t > 0$ and let $t^* \geq t.$ Define $G_0(s|W) = P(C \geq s|W)$ for any $s \geq 0$ and suppose that $G_0(T_i|W_i) \geq \epsilon$ 
almost surely for some $\epsilon > 0$.  Let $\Delta_i(t^*) = I(T_i(t^*) \leq C)$ and $T_i(t^*) = T_i \wedge t^*;$ 
in addition, define $\tilde{Z}_{im}(t) = I(\tilde{T}_i \leq t, M_i=m),$ $i=1,\ldots,n.$ 
Easy calculations then show 
\[
E\left[ \frac{\Delta_i(t^*)}{G_0(\tilde T_i(t^*)|W_i)} (\tilde{Z}_{im}(t) - \psi_m(t;W_i) ) ^2 \right] = E\left[ (Z_{im}(t) - \psi_m(t;W_i))^2 \right] = \Re(t,\psi_m)
\]
for a fixed $\psi_m(t;w).$
This risk equivalence motivates the construction of an IPCW-type observed data loss function that is an unbiased
estimator for $\Re(t,\psi_m).$
Define for any suitable survivor function $G(\cdot|\cdot)$ 
			\begin{gather}
      \label{eq:2}
			L_{m,t}^{ipcw}({\cal O},\psi_m; t^*, G)  =	\frac{1}{n}\sum_{i=1}^{n}\sum_{l=1}^L I\{W_i \in \mathcal{N}_l\}\bigg{[}\frac{\Delta_i(t^*)\{\tilde{Z}_{im}(t) - \beta_{lm}(t)\}^2}{G(\tilde{T}_i(t^*)|W_i)}\bigg{]};
			\end{gather}
then, $L_{m,t}^{ipcw}({\cal O},\psi_m; t^*, G_0)$ has the same risk as \eqref{eq:0} and		
it follows that \eqref{eq:2} is minimized by
\begin{gather}
\label{betaIPCW}
\hat{\beta}^{ipcw}_{lm}(t; t^*, G) = \frac{  \sum_{i=1}^n I\{W_i \in \mathcal{N}_l\} \frac{\Delta_i(t^*) \tilde{Z}_{im}(t)}{G(\tilde{T}_i(t^*)|W_i)}}
{\sum_{i=1}^n I\{W_i \in \mathcal{N}_l\} \frac{\Delta_i(t^*)}{G(\tilde{T}_i(t^*)|W_i)}}, l=1,\ldots,L.
\end{gather}
Observe that 
 \eqref{eq:2} and \eqref{betaIPCW}
respectively reduce to \eqref{eq:0} and \eqref{beta-full} when censoring is absent.

When $K=1$ and $t^* = \infty,$ we have
$\Delta_i(\infty) = \Delta_i$ and $T_i(\infty) = T_i;$ the loss
function \eqref{eq:2} is then just a special case of that considered in 
\citet{molinaro2004tree} (see also \citealp{steingrimsson2016doubly}).
Similarly, for $K=1$ and setting $t^* = t,$
the loss function \eqref{eq:2} is just that considered
in  \citet{lostritto2012partitioning}.
Hence,  \eqref{eq:2}  extends several existing
loss functions to the problem of estimating a CIF. In practice,
an estimator $\hat G(\cdot|\cdot)$ for $G_0(\cdot|\cdot)$ is used in \eqref{eq:2}; popular approaches here include product-limit estimators
derived from the Kaplan-Meier and Cox regression estimation procedures.

Given $t > 0,$ different choices of $t^* \geq $ in \eqref{eq:2} generate different losses, hence different CART algorithms. However, 
there are just two important choices of $t^* \geq t$ in \eqref{eq:2}: 
$t^* = \infty$ (standard IPCW; \textit{IPCW}$_{\!1}$) and $t^* = t$ (modified IPCW; \textit{IPCW}$_{\!2}$). In selecting $t^* = t,$ 
any observations that are censored after time $t$ can still contribute (all-cause failure) 
information to \eqref{eq:2};  as $t^* \rightarrow \infty$, the influence of these observations eventually vanishes. 
Consequently, the \textit{IPCW}$_{\!2}$ loss uses more of the observed data in calculating the loss function and associated minimizers compared to the \textit{IPCW}$_{\!1}$ loss and one can therefore expect a regression tree built using \eqref{eq:2} for $t^*=t$ to perform as well or better than one derived from \eqref{eq:2} with $t^*=\infty$. The use of $t^* = t$ in place of $t^* = \infty$ has an additional practical advantage: since $T_i(t^*) \leq T_i$ for every $i,$  we have $\hat G(\tilde T_i(t^*)|W_i) \geq \hat G(\tilde T_i|W_i),$ making it easier to 
satisfy the empirical positivity condition required for implementation.

\subsection{Improving the Efficiency of IPCW Loss Functions}
\label{CIF-obs3}

\subsubsection{Estimating $\Re(t,\psi_m)$ by augmenting the IPCW loss }
As in \cite{steingrimsson2016doubly}, one can employ semiparametric estimation theory for missing data to construct an improved estimator of  the full data risk $\Re(t,\psi_m).$ In particular, one can augment the IPCW loss function \eqref{eq:2} with additional information on
censored subjects, thereby incorporating additional information into the tree building process.

Consider first the loss function \eqref{eq:2} with $t^* = \infty;$ we denote this by $L_{m,t}^{ipcw}({\cal O},\psi_m;  G).$  
Define $\Psi_0 = \{ \psi_{0r}(s; w), s \geq 0, w \in {\cal S}, r=1,\ldots,K \}$ as the set of
CIFs of interest and let $\Psi$ denote a corresponding model that may or
may not coincide with $\Psi_0$.
For any $t, u \geq 0$ and $w \in {\cal S},$ define $V_{lm}(u;t,w,\Psi) = 
E_{\Psi}[(Z_{m}(t) - \beta_{lm}(t))^2|T \geq u,W= w];$ 
it is shown later how this expression depends on $\Psi.$
Then, fixing $\beta_{1m}(t),\ldots,\beta_{Lm}(t),$ the augmented estimator of 
$\Re(t,\psi_m)$ having the smallest variance
that can be constructed from the unbiased estimator $L_{m,t}^{ipcw}({\cal O},\psi_m;  G_0)$ is given by 
$L_{m,t}^{dr}({\cal O},\psi_m; G_0,\Psi_0) 
= L_{m,t}^{ipcw}({\cal O},\psi_m; G_0) 
+ L_{m,t}^{aug}({\cal O},\psi_m; G_0, \Psi_0)$ 
where,  for suitable models $\Psi$ and $G(\cdot|\cdot),$ 
\begin{gather}
	\label{eq:4a1}
L_{m,t}^{aug}(\mathcal{O},\psi_m; G, \Psi) =
 \frac{1}{n}\sum_{l=1}^{L}\sum_{i=1}^{n}I\{W_i \in \mathcal{N}_l\}
\int_0^{\tilde T_i} \frac{V_{lm}(u;t,W_i,\Psi)}{G(u|W_i)}dM_G(u|W_i)
\end{gather}
and 
$M_G(t|w) = I(\tilde{T} \leq t,\Delta = 0) - \int_0^t I(\tilde{T} \geq u)d\Lambda_G(u|w)$
with $\Lambda_G(t|w)$ denoting the cumulative hazard function corresponding 
to $G(t|w)$ \citep[cf.\ ][Sec.\ 9.3, 10.3]{tsiatis2007semiparametric}.
The ``doubly robust'' loss function
$L_{m,t}^{dr}({\cal O},\psi_m; G,\Psi)$ reduces to a special case of the class of 
loss functions proposed in \citet{steingrimsson2016doubly} when $K=1.$

Instead of augmenting $L_{m,t}^{ipcw}({\cal O},\psi_m;  G)$ (i.e., \textit{IPCW}$_{\! 1}$), one might instead 
augment  \eqref{eq:2} when $t^* = t$ (i.e., \textit{IPCW}$_{\! 2}$). However, it turns out that the resulting  loss
function is identical to $L_{m,t}^{dr}({\cal O},\psi_m; G,\Psi);$ see the Appendix (Section \ref{loss-equiv-dr}). 
Hence, no additional efficiency is gained from augmenting  \textit{IPCW}$_{\! 2}$ and 
it suffices to consider $L_{m,t}^{dr}({\cal O},\psi_m; G,\Psi)$ only.

Returning to $L_{m,t}^{dr}({\cal O},\psi_m; G,\Psi):$  because $Z_{m}(t)$ is binary, we have
\begin{equation}
\label{simple-V}
V_{lm}(u;t,w,\Psi) = y_{m}(u;t,w,\Psi) - 2 y_{m}(u;t,w,\Psi) \beta_{lm}(t) + \beta^2_{lm}(t)
\end{equation}
for any suitable $\Psi$ (e.g., $\Psi_0$), where $y_{m}(u;t,w,\Psi) = E_{\Psi}\{Z_{m}(t)|T \geq u, W=w\}$
reduces to
\begin{gather}
\label{eq:2b}
		y_{m}(u;t,w,\Psi) =
			\begin{cases}
			\dfrac{P_{\Psi}(u \leq T \leq t, M=m|W=w)}{P_{\Psi}(T \geq u | W=w)} & \text{if} \quad u \leq t \\
			0 & \text{otherwise}
			\end{cases}.
\end{gather}
The notation $E_{\Psi}$ and $P_{\Psi}$ means that these quantities are calculated under the CIF
model specification $\Psi$. Hence, under a model $\Psi$, the calculation of $L_{m,t}^{dr}({\cal O},\psi_m; G,\Psi)$
requires estimating both the CIF for cause $m$ and the  all-cause probability $P_{\Psi}(T \geq u| W= w).$

Considering $L_{m,t}^{dr}({\cal O},\psi_m; G,\Psi)$  as a function of the $L$ scalar parameters 
$\beta_{1m}(t),\ldots,\beta_{Lm}(t)$ only and differentiating with respect to each one, it can be shown that
\begin{gather}
\label{eq:5}
\tilde{\beta}_{lm}^{dr}(t; G,\Psi) = \dfrac{\displaystyle \sum_{i=1}^n I\{W_i \in \mathcal{N}_l\}[\widetilde{TS}_{1,im}^1(t) + \widetilde{TS}_{2,im}^1(t) ] }{\displaystyle \sum_{i=1}^n I\{W_i \in \mathcal{N}_l\} [\widetilde{TS}_{1,im}^0 + \widetilde{TS}_{2,im}^0 ]}, ~l=1,\ldots,L
			\end{gather}
minimize  $L_{m,t}^{dr}({\cal O},\psi_m; G,\Psi),$ 
where 
			\begin{gather*}
		\widetilde{TS}_{1,im}^0(t)  = \frac{\Delta_i}{G(\tilde{T}_i | W_i)} \quad \quad
			\widetilde{TS}_{2,im}^0(t)  = \int_0^{\tilde{T}_i} \frac{dM_G(u|W_i)}{G(u|W_i)} \\
			\widetilde{TS}_{1,im}^1(t)  = \frac{\tilde{Z}_{im}(t)\Delta_i}{G(\tilde{T}_i| W_i)} \quad \quad 
			\widetilde{TS}_{2,im}^1(t)  = \int_0^{\tilde{T}_i} \frac{y_m(u;t,W_i,\Psi)}{G(u|W_i)}dM_G(u|W_i).
			\end{gather*}
The validity of this result relies on the assumption that $G(\tilde{T}_i|W_i) \geq \epsilon > 0$ for
some $\epsilon$ and each $i=1,\ldots,n.$ Under this same assumption, Lemma 1 of \cite{strawderman2000estimating}
implies
\begin{gather*}
		\widetilde{TS}_{1,im}^0  + \widetilde{TS}_{2,im}^0 =
		\dfrac{\Delta_i}{G(\tilde{T}_i | W_i)} + \dfrac{1-\Delta_i}{G(\tilde{T}_i|W_i)} - \displaystyle \int_{0}^{\tilde{T}_i}\dfrac{d \Lambda_G(u|W_i)}{G(u|W_i)} = 1;
\end{gather*}
letting $N_l = \sum_{i=1}^n I\{W_i \in \mathcal{N}_l\}, ~l=1,\ldots,L,$
it follows that \eqref{eq:5} can be rewritten as
\begin{gather}
		\label{eq:5a}
	\hat{\beta}_{lm}^{dr}(t;G,\Psi) =  \frac{1}{N_l}  \sum_{i=1}^n I\{W_i \in \mathcal{N}_l\}[\widetilde{TS}_{1,im}^1(t)  + \widetilde{TS}_{2,im}^1(t) ] , ~l=1,\ldots,L.
\end{gather}
As in  Section \ref{CIF-obs},
$L_{m,t}^{dr}({\cal O},\psi_m; G,\Psi)$ 
and \eqref{eq:5a} respectively reduce to \eqref{eq:0} and \eqref{beta-full} when censoring is absent.

The specification $\tilde G(t|w) = 1$ for all $t \geq 0$ and 
$w \in {\cal S}$ generates an interesting special case of $L_{m,t}^{dr}({\cal O},\psi_m; \tilde G,\Psi)$
despite the fact that $\tilde G(\cdot|\cdot)$ is necessarily incorrectly modeled in the presence of censoring.
In particular, for any suitable choice of $\Psi$,  
(i) $L_{m,t}^{dr}({\cal O},\psi_m; \tilde G,\Psi) = \sum_{l=1}^L L_{ml,t}^{bj}({\cal O},\psi_m; \Psi)$ where 
\[
L_{ml,t}^{bj}({\cal O},\psi_m; \Psi)  =  
  \frac{1}{n} \sum_{i=1}^n 
I\{W_i \in \mathcal{N}_l\} [\Delta_i\{\tilde Z_{im}(t) - \beta_{lm}(t)\}^2 + (1-\Delta_i)V_{lm}(\tilde{T}_i;t,W_i,
\Psi)];
\]
and, (ii) for $\Psi = \Psi_0,$ $L_{m,t}^{dr}({\cal O},\psi_m; \tilde G,\Psi_0)$ is an
unbiased estimator of the risk  $\Re(t,\psi_m)$.
In fact, noting that \eqref{simple-V} implies $V_{lm}(\tilde{T}_i;t,W_i,
\Psi)$ can be rewritten in terms of $y_m(\tilde{T}_i; t,w, \Psi)$
for every $i,$ the minimizer of $L_{ml,t}^{bj}({\cal O},\psi_m; \Psi)$ is given by
\begin{gather*}
\tilde{\beta}_{lm}^{bj}(t;\Psi) = 
\frac{1}{N_l} 
 \sum_{i=1}^n I\{W_i \in \mathcal{N}_l\} [\Delta_i \tilde{Z}_{im}(t) + (1-\Delta_i)y_m(\tilde{T}_i; t,W_i,\Psi)].
\end{gather*}
That is, under the loss $L_{ml,t}^{bj}({\cal O},\psi_m; \Psi),$
the estimator for $\beta_{lm}(t)$ is the Buckley-James (BJ) estimator 
 of the mean response within node ${\cal N}_l$
 \citep{buckley1979linear},
 an estimator that can also be derived directly from \eqref{eq:5a} by setting $G = \tilde G.$
For this reason, we refer to $L_{m,t}^{dr}({\cal O},\psi_m; \tilde G,\Psi)$ as the Buckley-James loss function.

\subsubsection{Composite augmented loss functions with multiple time points}
\label{composite loss}

Under a tree model of the form \eqref{psi mod}, and similarly to methods used for survival trees 
\citep[e.g.,][]{leblanc1992relative, molinaro2004tree, steingrimsson2016doubly}, 
the quantity being estimated within each terminal node depends on the choice of $t$ but the underlying partition structure is time-invariant. That is, the effects of baseline covariates and their interactions on the CIF are not allowed to change over time. As a result, for a given $m$, we can
further reduce variability when estimating $\psi_{0m}(t|w)$ by considering loss functions constructed from $L_{m,t}^{dr}({\cal O},\psi_m; G,\Psi)$ that incorporate information over several time points. 

Recall that
$L_{m,t}^{dr}({\cal O},\psi_m; G,\Psi) = L_{m,t}^{ipcw}({\cal O},\psi_m; G) 
+ L_{m,t}^{aug}({\cal O},\psi_m; G, \Psi)$
where $L_{m,t}^{ipcw}({\cal O},\psi_m; G)$ is given by \eqref{eq:2} (i.e., for
$t^* = \infty$) and 
$L_{m,t}^{aug}({\cal O},\psi_m; G, \Psi)$ is given by \eqref{eq:4a1}.
For a given set of time points $0 < t_1 < t_2 \ldots < t_J < \infty,$ a simple composite loss function 
for a given event type $m$ can be formed by calculating 
\begin{gather}
\label{eq:7a}
L_{m,\mathbf{t}}^{mult,dr}(\mathcal{O},\psi_m;G,\Psi) = \sum_{j=1}^J w_j L_{m,t_j}^{dr}({\cal O},\psi_m; G,\Psi),
\end{gather}
where $w_1,\ldots,w_J$ are positive weights that, without loss of generality, can be assumed to sum to one.  
Minimizing \eqref{eq:7a} with respect to $\beta_{lm}(t_j), j=1,\ldots,J; l = 1,\ldots, L,$  gives
\begin{gather}
\label{eq:7b}
		\tilde{\beta}_{lm}^{mult,dr}(t_j; G,\Psi) = \frac{1}{N_l} 
\ \sum_{i=1}^{n} I\{W_i \in \mathcal{N}_l\}[\widetilde{TS}_{1,im}^1(t_j) + \widetilde{TS}_{2,im}^1(t_j)].
\end{gather}
The terminal node estimators \eqref{eq:7b} are exactly equivalent to (\ref{eq:5a}) computed for $t=t_j$ and do not depend on $w_1,\ldots,w_J$. However, the corresponding assessment of total loss and thus decisions to split nodes at any given cutpoint for any given covariate 
are directly influenced by the combined specifications of $(t_j,w_j), j = 1,\ldots,J.$  Consequently, performance
gains in the resulting tree-based estimates for $\psi_{m0}(t_j;w), j=1,\ldots,J$
occur in the determination of when and where to split, rather than in the manner by
which individual terminal nodes are estimated. In the absence of censoring, the indicated composite loss function
also reduces to that which would be computed by extending the loss function introduced in Section
\ref{CIF-full} in the manner described above.

\section{Simulation Studies}
\label{sims}
In this section, we will summarize the structure and results of several simulation studies designed
to study the performance of regression trees built using the IPCW, Buckley James, and doubly robust Brier loss functions
introduced in the previous section. We first describe how data are generated. Several performance evaluation measures are used to 
evaluate tree-building performance; these are described next, followed by a summary of the results.
 
\subsection{Simulation Settings}
\label{sim set}
\begin{figure}[!htbp]
\caption{Three settings in simulation 1. Black line : $\psi_{01}(\cdot|Z(\boldsymbol{W})=1)$, 
Blue line : $\psi_{01}(\cdot|Z(\boldsymbol{W})=0)$, Red line : $\psi_{02}(\cdot|Z(\boldsymbol{W})=1)$, Green line : $\psi_{02}(\cdot|Z(\boldsymbol{W})=0)$.}
\includegraphics{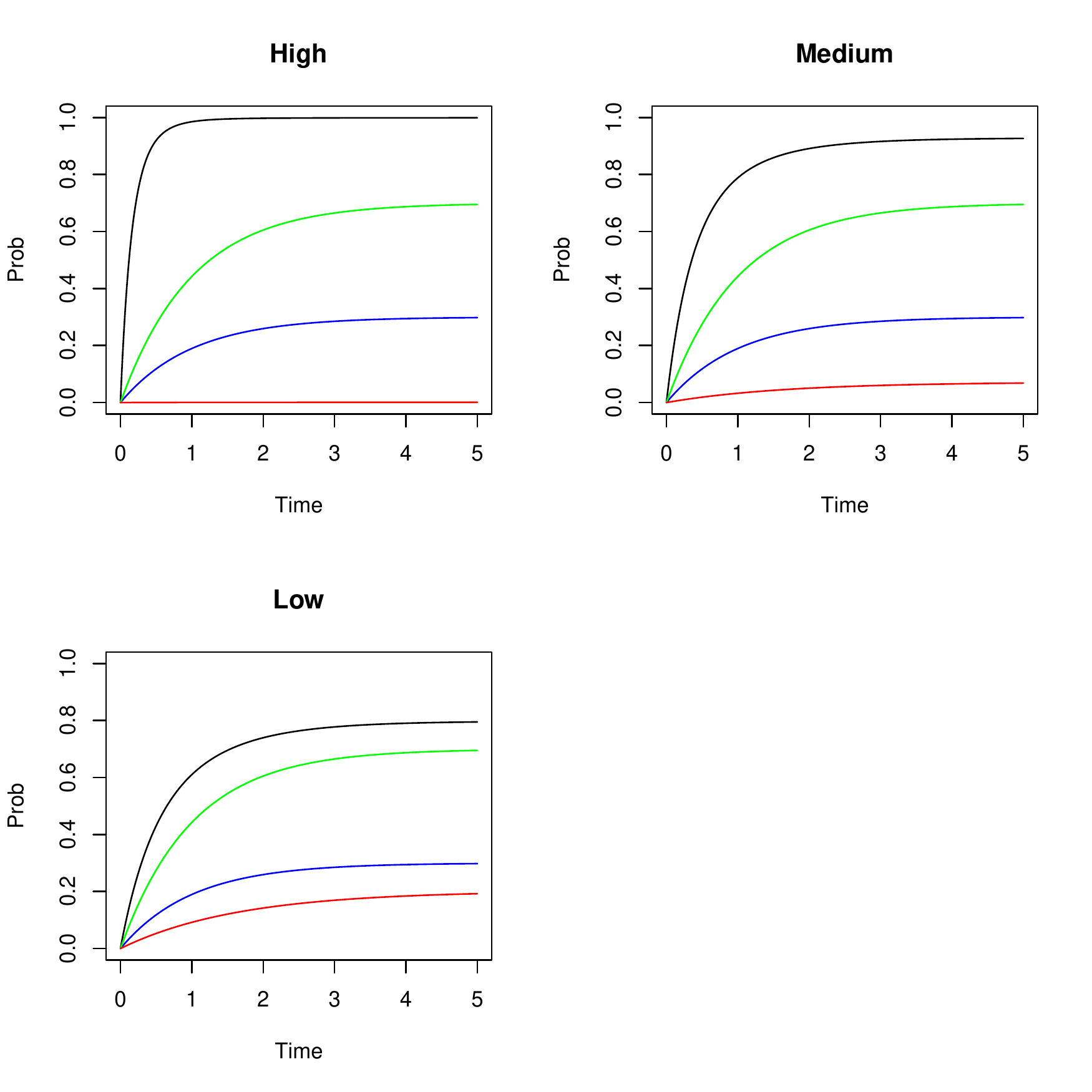}
\label{fig1}
\end{figure}

Covariates $W_1,W_2,W_3,W_4,W_5,W_6,W_7,W_8,W_9,W_{10} \sim Unif(0,1)$
are generated independently of each other; let 
$\boldsymbol{W} = (W_1,W_2,W_3,W_4,W_5,W_6,W_7,W_8,W_9,W_{10})'.$
The underlying competing risks model assumes $K=2$ and takes the form
				\begin{gather}
				\label{true sim CIF}
				\psi_{01}(t;\boldsymbol{W}) = 1 - (1-p(1-e^{-t}))^{\exp{(\beta_{1} Z(\boldsymbol{W}) )}} \\
				\psi_{02}(t;\boldsymbol{W}) = (1-p)^{\exp{(\beta_{1} Z(\boldsymbol{W}))}} \times (1-\exp{(-t\exp{(\beta_{2} Z(\boldsymbol{W}))})})
				\end{gather} 
where $Z(\boldsymbol{W}) = I(W_1 \leq 0.5 \ \& \ W_2 > 0.5)$ and 
$\beta_{1}$ and $\beta_{2}$ are regression coefficients \citep[cf.,][]{fine1999proportional}.
Note that this formulation satisfies the additivity constraint
$\psi_{01}(\infty;\boldsymbol{W}) + \psi_{02}(\infty;\boldsymbol{W}) =1.$
Three settings are considered: (i) $\beta_{1} = 3$ and $\beta_{2} = -0.5$ with $p=0.3$ (High Signal); (ii) $\beta_{1} = 2$ and $\beta_{2} = -0.5$ with $p=0.3$ (Medium Signal); and, (iii) $\beta_{10} = 1.5$ and $\beta_{20} = -0.5$ with $p=0.3$ (Low Signal). 
The corresponding cumulative incidence functions for these settings are shown in Figure \ref{fig1}.


Similarly to \cite{steingrimsson2016doubly}, we respectively generate training sets of size 250, 500 and 1000 
(with noninformative censoring) for the purposes of estimation and an independent uncensored test set of size 2000 for 
evaluating performance. Censoring is exponentially distributed with a rate $\gamma$ chosen to give approximately 50\% censoring 
on $T$. For each training set size, 500 simulations are run for each of the three settings.
 
\subsection{Performance Measures}
\label{perf}

For assessing estimator performance, it is assumed there is an underlying true tree structure to be recovered. Let $\hat{\psi}_1(t|W_i)$ be
the tree-predicted cumulative incidence for the cause of interest for a subject having covariates $W_i$; let $\psi_{01}(t|W_i)$ be the corresponding true cumulative incidence function. 
Then, as a measure of  predictive performance, we consider 
		\begin{gather}
		\label{eq:17}
		\frac{1}{n_{test}} \sum_{i=1}^{n_{test}} \{ \hat{\psi}_1(t|W_i) - \psi_{01}(t|W_i)\}^2
		\end{gather}
where  $n_{test}$ is the number of observations in the test dataset. 
We also use several other performance measures focused more on the identification of the underlying tree structure:
\begin{itemize}
\item $|$fitted size$-$true size$|$ : This quantity measures difference between the size of fitted tree and the size of the true tree, where size denotes the number of terminal nodes. It can be seen from Figure \ref{fig1} that the true tree size is 3.
\item NSP: This quantity is defined as $(n_{sim})^{-1} \sum_{i=1}^{n_{sim}} NS_i$ where $NS_i$ represents number of times any of the covariates $W_3,\ldots, W_{10}$ (i.e., covariates that do not affect the true CIF) appear in simulation run $i$.
\item PCSP : This quantity is defined as $(n_{sim})^{-1} \sum_{i=1}^{n_{sim}} CT_i$ where $CT_i = 1$ if the fitted tree in simulation $i$ has the same covariates {and the same number of splits} as the true tree that defines the CIF; otherwise $CT_i = 0$. While it can happen that a fitted tree may involve only the covariates $W_1$ and $W_2$, it cannot be expected to have exactly the same splits as the true tree 
(especially for a continuous covariate); thus, in this measure, a `correct tree' 
is one that looks like the true tree but may involve different split points.
\end{itemize}

%

\subsection{Estimation of $G_0$ and $\Psi_0$}
\label{nuisance}

The IPCW and doubly robust Brier loss functions require estimation of $G_0(\cdot|\cdot).$ The Buckley-James
and doubly robust Brier loss functions both rely on the function 
 $y_m(\cdot;t, w, \Psi)$ in \eqref{eq:2b}, the optimal choice requiring specification of both $\psi_{0m}(u|w)$ 
 and the event-free survival function $\bar \psi_0(u|w) = P_{\Psi_0}(T > u | w)$ for $u > 0, w \in {\cal S}$.

Estimation of $G_0(\cdot|\cdot)$ is comparatively straightforward; for example, one may use
regression procedures (e.g., Cox regression models, random survival forests) or
product-limit estimators as appropriate. For all simulations considered here, the censoring distribution $G_0(\cdot|\cdot)$ 
is estimated by the Kaplan-Meier method.  
For building an IPCW-type tree with $t^* = \infty$ in \eqref{eq:2}, the resulting censoring weight
may not satisfy the required positivity assumption. Hence, similarly to \cite{steingrimsson2016doubly}, we replace $t^* = \infty$ by 
$t^* = \tau$ when calculating \textit{IPCW}$_{\!1},$ where  $\hat{G}(\tilde T_i(\tau)) \geq 0.05, i=1,\ldots,n$ 
(i.e., the marginal truncation rate of observed survival times is 5\%). 
No such modification is required when calculating \textit{IPCW}$_{\!2},$ that is, when computing 
\eqref{eq:2} with $t^* = t = t_j, j =1, 2, 3.$

Calculation of the Buckley-James
and doubly robust Brier loss functions in practice requires using an estimated model $\hat \Psi$
in place of $\Psi_0$ when computing \eqref{eq:2b}. Parametric models have been proposed that ensure the 
compatibility between the models used for  $\psi_{m}(\cdot|\cdot)$ and $\bar \psi(\cdot|\cdot)$
as well as the ability to calculate probabilities for every $u > 0;$ see, for example, 
\citet{jeong2006direct}, \citet{jeong2007parametric}, \citet{cheng2009modeling}, and \citet{shi2013constrained}.
Compared to existing semiparametric methods (e.g., \citealp{fine1999proportional}; \citealp{scheike2008predicting}; \citealp{eriksson2015proportional}), the parametric likelihood methods of \cite{jeong2006direct} and \cite{jeong2007parametric} enable researchers to estimate two or more CIFs jointly. However,  \cite{cheng2009modeling} and \cite{shi2013constrained} show that even these methods fail to enforce the additivity constraint $\sum_{m=1}^M P_{\Psi}(T \leq \infty, M=m;W=w) = 1$ and propose methods that resolve this inconsistency.

Since our focus is on a particular cause $m,$ we will without loss of generality assume that 
$m=1$ and that $K=2$ (i.e., $m=2$ captures all causes $m \neq 1$); in this case, 
specification
of $\psi_{1}(\cdot|\cdot)$ and $\bar \psi(\cdot|\cdot)$ is equivalent to specifying 
$\Psi = (\psi_{1}(\cdot|\cdot), \psi_{2}(\cdot|\cdot)).$ Our simulation study
considers the following estimators $\hat \Psi$ when $K=2:$ \\
\begin{enumerate}
\item \textit{RF(npar)} :  random survival forests as proposed in \citet{ishwaran2014random}.

\item \textit{RF+lgtc}: a hybrid method of random survival forests and parametric modeling; 

\item \textit{FG(true)}: the simulation model 
of Section \ref{sim set} is an example of the class of models considered in 
\citet{jeong2007parametric}. For $m=1,2,$ $\psi_{m}(\cdot|\cdot)$ are estimated
by maximum likelihood under a correctly specified parametric model. 
  
\item \textit{lgtc+godds}: for $m=1,2,$ $\psi_{m}(\cdot|\cdot)$ is estimated by maximum likelihood
using the parametric cumulative incidence model proposed in \cite{shi2013constrained}.
\end{enumerate}
Further details on the \textit{RF(npar)} and \textit{RF+lgtc} estimators,
including the parametric models used in \#2 and \#4 above, 
may be found in Section \ref{sec:est-CE} of the Appendix.
We anticipate that \textit{FG(true)} will lead to the best performance 
because the underlying models for the event of interest and for censoring are both specified correctly.
As in \cite{steingrimsson2016doubly},  the estimates  $\hat G(\cdot|\cdot)$ and $\hat \Psi$ depend only on ${\cal O}$ and
are computed in advance of
the process of building the desired regression tree for estimating $\psi_{01}(\cdot|\cdot).$

\subsection{Simulation Results} 
\label{main sim res}

Each combination of loss function and method for estimating $G_0$ and/or $\Psi_0$ leads to a distinct
algorithm. In this section,  we show results for the IPCW-type estimators with $t^*= t_j, j=1,2,3$ and $t^* = \tau$; the
Buckley James Brier loss, where  \textit{RF(npar)}  (algorithm \textit{BJ-RF (npar)}) and  \textit{FG(true)} (algorithm \textit{BJ-FG (true)}) 
are respectively used to estimate $\Psi_0$;  and, the doubly robust Brier loss function, where
\textit{RF(npar)}  (algorithm \textit{DR-RF (npar)}) and  \textit{FG(true)} (algorithm \textit{DR-FG (true)}) 
are respectively used to estimate $\Psi_0$. Results using \textit{RF+lgtc} and \textit{lgtc+godds}
for estimating $\Psi_0$ are summarized in the Appendix (Section \ref{add sim res}).

\begin{figure}[!htbp]
\caption{Prediction error (multiplied by 100) for event 1 
using composite loss for $n=500$. Rows represent signal strength; columns indicate 
time values considered. \textit{IPCW}$_{1}$ and \textit{IPCW}$_2$ are standard and modified
IPCW, respectively; \textit{BJ-RF(npar)} and \textit{BJ-FG(true)} use 
Buckley-James loss functions with the augmentation term estimated via 
nonparametric random forests and under the correct simulation model, respectively; 
and, \textit{DR-RF(npar)} and \textit{DR-FG(true)} use the doubly robust loss function
with the augmentation term estimated via 
nonparametric random forests and under the correct simulation model, respectively.}
 \begin{minipage}[b][1\textheight][s]{1.05\textwidth}
  \centering
	  \includegraphics[height=0.3\textheight,width=\textwidth]{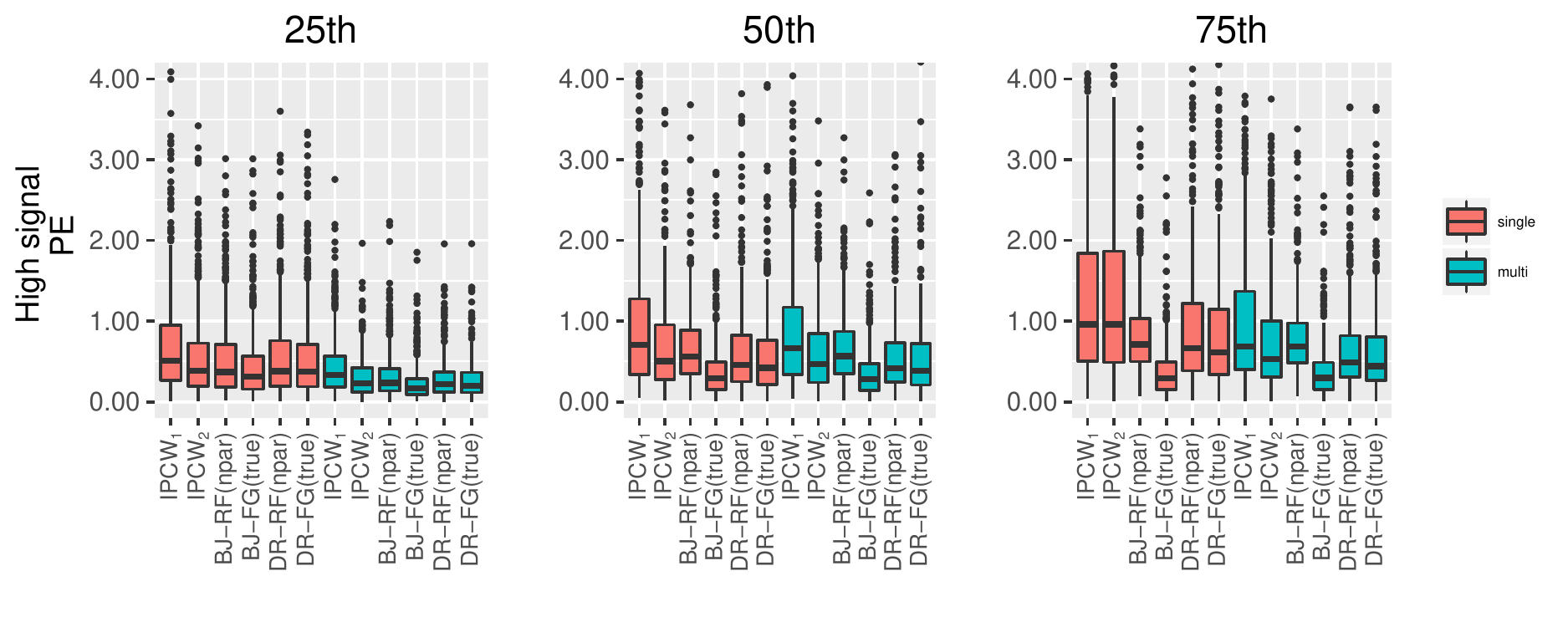}
  \vspace{-0.6cm}
  \includegraphics[height=0.3\textheight,width=\textwidth]{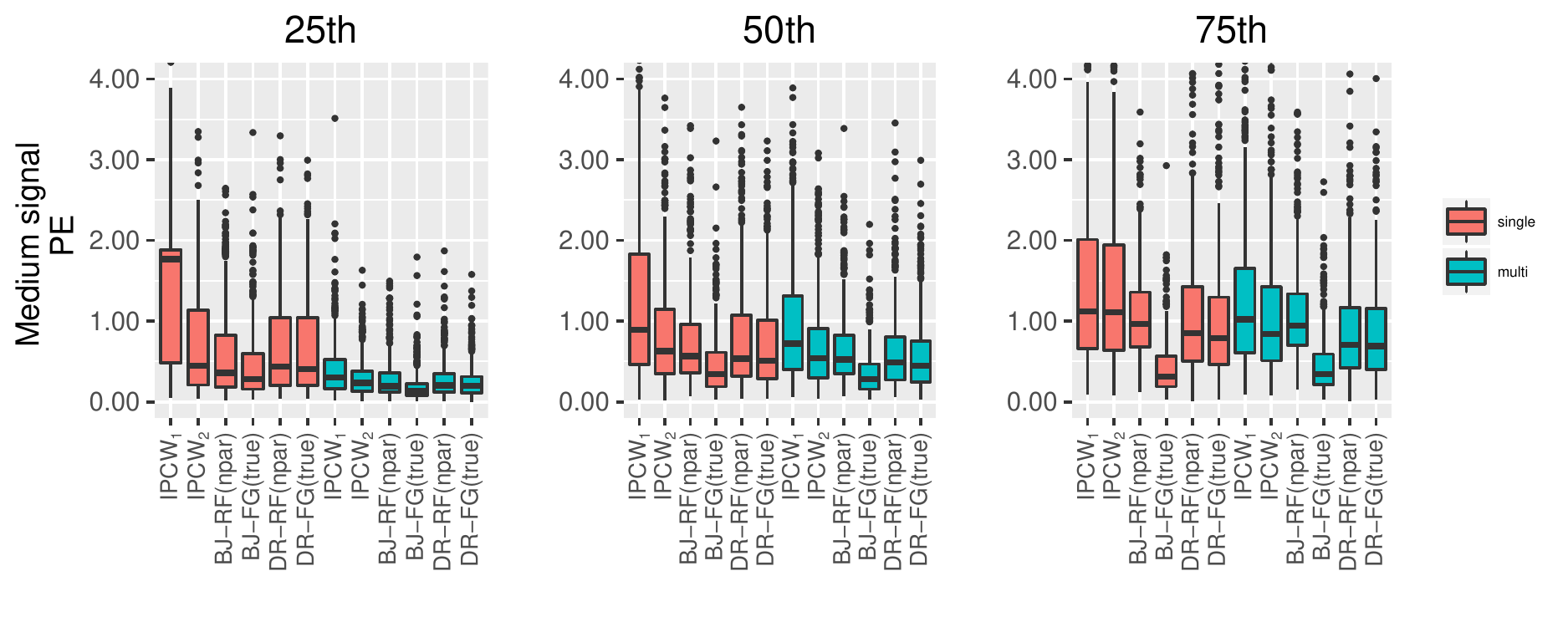}
  \vspace{-0.6cm}
  \includegraphics[height=0.3\textheight,width=\textwidth]{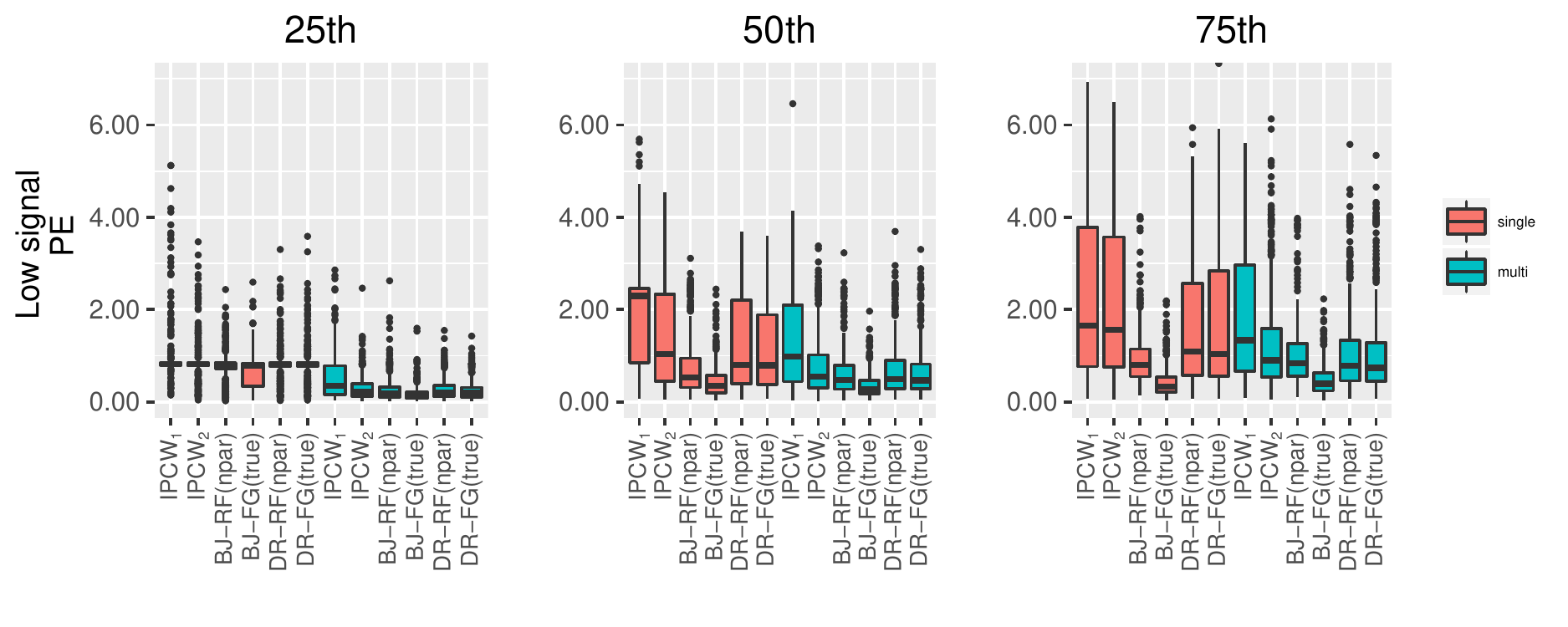}
\end{minipage}
  \vspace{-3.75cm}
\label{fig2}
\end{figure}

In fitting the trees, we consider three fixed time points $t_1$, $t_2$ and $t_3$, respectively representing the 
25th, 50th and 75th quantile of the true marginal distribution of $T$. These times are used for building individual 
trees as well as for composite loss functions
with the respective training set sizes $n=250, 500$ and $1000$. Below, we focus on the results 
with $n=500$ using composite loss functions as described in Section \ref{composite loss}.
Table \ref{tab1} summarizes the tree fitting performance measures 
for estimating $\psi_{01}(\cdot|\cdot)$ in \eqref{true sim CIF} for $n=500$ using a composite loss 
function having weights $w_1 = w_2 = w_3 = 1/3$. 
%
%
For the chosen metrics, and irrespective of signal strength, there is a clear benefit to using augmented loss functions over IPCW loss functions, 
the exception being \textit{BJ-RF (npar)} (i.e., the Buckley-James loss with $\hat \Psi$ estimated by random survival forests). 
Overall, the \textit{BJ-FG (true)} algorithm that uses the correct model for $\Psi_0$  performs best, followed by 
\textit{DR-FG (true)}, \textit{DR-RF (npar)}, \textit{IPCW}$_{\! 2},$ \textit{BJ-RF (npar)} and then \textit{IPCW}$_{\! 1}$.  
Prediction error performance is considered in Figure \ref{fig2}, which shows boxplots of the mean squared prediction error given in \eqref{eq:17}
under equally weighted composite loss.  Here, the \textit{BJ-FG (true)} algorithm again performs uniformly best, with the augmented methods
\textit{DR-FG (true)}, \textit{DR-RF (npar)}, and \textit{BJ-RF (npar)} all performing similarly, followed by 
\textit{IPCW}$_{\! 2}$ and then \textit{IPCW}$_{\! 1}.$

In general, the methods that incorporate information beyond that used by \textit{IPCW}$_1$ exhibit significant
improvement in performance, with the methodology for estimating quantities needed to compute the augmented 
loss function playing a smaller role, particularly as sample size increases.  Considering only methods where no knowledge of 
$\Psi_0$ is used, the doubly robust loss function also provides gains over both the Buckley-James and \textit{IPCW}$_{\! 2}$ 
loss functions; however, the gains achieved are substantially less compared to the gains over  \textit{IPCW}$_{\! 1}$.  This  phenomenon 
is expected and can be explained by noting that (i) the censoring distribution is being consistently estimated in all cases; 
(ii) \textit{IPCW}$_{\! 2}$ can be viewed as an augmented version of \textit{IPCW}$_{\! 1};$ and, 
(iii) \textit{IPCW}$_{\! 2}$ only requires estimating one infinite dimensional parameter.

Numerical and graphical results for other methods of nuisance parameter estimation and other settings (a single time point loss and composite loss with various training set sample sizes) are shown in the Appendix. These results demonstrate no significant deviations from the trends and
observations summarized above.  Some important conclusions that can be drawn from the combined set of results include the
following: (i) the method for estimating $\Psi$ in \eqref{eq:2b}  has little overall impact on performance, with greater differences arising
from the type of loss (IPCW versus DR versus BJ) and whether or not a composite loss function is used; (ii)  getting the data generating model exactly right yields noticeable gains, and there is otherwise a degree of robustness across methods, particularly for doubly robust loss; and, (iii) the performance of IPCW$_2$ is often reasonably close to that for the augmented losses, particularly with composite loss, and the
comparative degree of simplicity involved in implementing this method has much to recommend it.


\begin{table}[!htbp]
\centering
\caption{Numerical summaries for trees built using composite loss
for $n=500$. \textit{IPCW}$_{1}$ and \textit{IPCW}$_2$ are standard and modified
IPCW, respectively; \textit{BJ-RF(npar)} and \textit{BJ-FG(true)} use 
Buckley-James loss functions with the augmentation term estimated via 
nonparametric random forests and under the correct simulation model, respectively; 
and, \textit{DR-RF(npar)} and \textit{DR-FG(true)} use the doubly robust loss function
with the augmentation term estimated via 
nonparametric random forests and under the correct simulation model, respectively.}
\resizebox{\columnwidth}{!}{%
\begin{tabular}{crrrrrrrrr}
  \hline
 & \multicolumn{3}{c}{High Sig} & \multicolumn{3}{c}{Med Sig} & \multicolumn{3}{c}{Low Sig} \\ 
& $|$fitted-3$|$ & NSP & PCSP & $|$fitted-3$|$ & NSP & PCSP & $|$fitted-3$|$ & NSP & PCSP \\ 
  \hline 
 \textit{IPCW}$_1$ & 0.132 & 0.084 & 0.916 & 0.182 & 0.132 & 0.874 & 0.558 & 0.164 & 0.658 \\ 
 \textit{IPCW}$_2$ & 0.124 & 0.082 & 0.932 & 0.138 & 0.100 & 0.906 & 0.282 & 0.136 & 0.830 \\ 
 \textit{BJ-RF(npar)}  & 0.142 & 0.090 & 0.904 & 0.148 & 0.088 & 0.908 & 0.226 & 0.128 & 0.878 \\ 
 \textit{BJ-FG(true)}  & 0.092 & 0.068 & 0.940 & 0.102 & 0.064 & 0.934 & 0.118 & 0.082 & 0.918 \\ 
 \textit{DR-RF(npar)} & 0.066 & 0.052 & 0.958 & 0.092 & 0.056 & 0.940 & 0.216 & 0.118 & 0.856 \\ 
 \textit{DR-FG(true)}  & 0.058 & 0.044 & 0.966 & 0.092 & 0.058 & 0.942 & 0.176 & 0.076 & 0.874 \\ \hline
\end{tabular}	
}
\label{tab1}
\end{table}

\section{Example: Lung Cancer Treatment Trial}
\label{RTOG9410}

We illustrate our methods using data from the RTOG 9410, randomized trial of patients with locally advanced inoperable 
non-small cell lung cancer.  The motivation for this trial was to ascertain whether sequential or concurrent delivery of chemotherapy and thoracic radiotherapy 
(TRT) is a better treatment strategy. 
The original RTOG 9410 study randomized 610 patients to three treatment arms: sequential chemotherapy followed by radiotherapy (RX=1);
once-daily chemotherapy concurrent with radiotherapy (RX=2); and, twice-daily chemotherapy concurrent with radiotherapy (RX=3).
The primary endpoint of interest was overall survival and the main trial analysis results were published in \cite{curran2011sequential}, demonstrating a survival benefit of concurrent delivery of chemotherapy and TRT compared with sequential delivery. Secondary analyses of the data using the time from randomization to the first occurrence of three possible outcomes are considered: in-field failure (cancer recurrence within the treatment field for TRT); 
out-field failure (cancer recurrence and distant metastasis outside of the treatment field for TRT); and, 
death without documented in-field or out-field failure (i.e., cancer progression). 
Among these event types, those that first experienced out-field failures are of particular interest
since these patients typically have suboptimal prognosis and may be candidates for more intensified 
treatment regimens intended to prevent distant metastasis, including but not limited to consolidative 
chemotherapy, prophylactic cranial irradiation (for brain metastases), and so on.
As such, patients that experienced both in-field failure and out-field failure were 
considered to be out-field failures for purposes of this analysis.
%

At the time the study database was last updated in 2009, there were 577 patients, with approximately 3\% censoring on the aforementioned outcomes. Because this censoring rate is so low, we removed these censored observations and compare the results of analyses of the 
resulting uncensored dataset (554 patients) to analyses of data created using an artificially induced censoring mechanism.   The purpose of doing this analysis is to evaluate how censoring affects the analysis and, in particular, illustrates how well the various procedures are able to recover the desired tree(s) that would be built had all outcomes been fully observed.  
We focus on building trees for each outcome using a composite loss function with 5 time points (5.2, 6.2, 8.5, 11.8, 19.0 months), selected as the 25th, 35th, 50th, 65th and 80th percentiles of the observed ``all cause" event time (i.e., $T$). Baseline covariates included in this analysis are Treatment (RX), 
Age, Stage (American Joint Committee on Cancer [AJCC] stage IIIB vs.\  IIIA or II), 
Gender, Karnofsky score (70, 80, 90 or 100), Race (White vs.\ non-White), and Histology (Squamous vs.\ non-Squamous).
Censoring is created according to a Uniform $[0, 50]$ distribution, generating approximately 29\% censoring on $T$. In addition to building a tree using the uncensored dataset (i.e., see Section \ref{composite loss}), we consider the methods \textit{IPCW}$_{\!1}$, \textit{IPCW}$_{\!2}$, 
\textit{BJ-RF(npar)}, \textit{BJ-RF(npar)}, \textit{DR-RF+lgtc}, \textit{DR-RF(npar)} and \textit{DR-RF+lgtc}; this will allow us to evaluate the effect of censoring as well as different models used in constructing the augmented loss. Similarly to \cite{steingrimsson2016doubly},  we use repeated 10-fold cross-validation to improve the stability of the tree building process and select the final tree using that having the lowest mean cross-validated error.  
For the in-field failure outcome, all methods (including those applied to the uncensored dataset)  lead to trees that consist only of a root node. Hence, we focus on outfield failure and death without progression below.

\begin{figure}[!ht]
\caption{Risk stratification for out-field failure. 
Panels contain estimated CIFs obtained using the following composite loss
functions: \textit{Uncensored} [top left]; \textit{DR-RF(npar)} [top right]; \textit{IPCW$_2$} [bottom left]; 
\textit{BJ-RF(npar)} [bottom right]. CIF estimates are calculated
at the following percentiles of observed all-cause event time:
15, 25, 35, 50, 65, 75, 80. Estimates are otherwise interpolated.}
\centering
\includegraphics[scale=0.9]{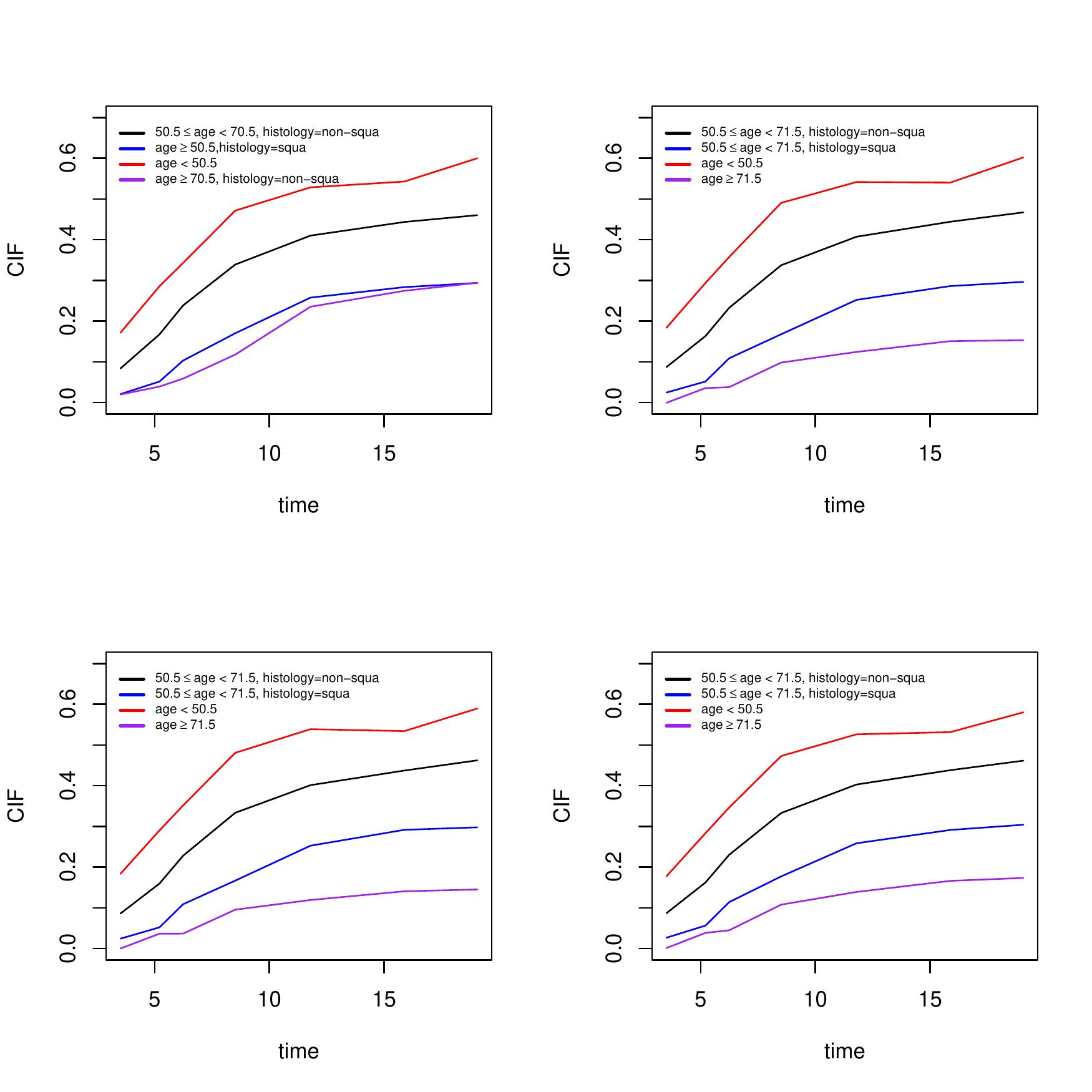}
\label{fig4}
\end{figure}

For out-field failure, the estimated tree obtained using the uncensored data creates 4 risk groupings (see Figure \ref{fig-of-u} in the Appendix): 
age $<$ 50.5; 50.5 $\leq$ age $<$ 70.5 with squamous histology; age $\geq$ 50.5 with squamous histology; and, age $\geq$ 70.5 with non-squamous histology. The plot of the CIF indicates that age $\geq$ 70.5 with non-squamous histology has the lowest risk, and that this risk is very similar
to those aged $\geq$ 50.5 with squamous histology.
With censored data, the \textit{IPCW}$_{\!1}$, \textit{IPCW}$_{\!2}$, \textit{BJ-RF(npar)} and \textit{DR-RF(npar)} methods all lead to the same tree structure; see 
%
%
Figures \ref{fig-of-outcen-ipcw1} -- \ref{fig-of-outcen-dr-npar} in the Appendix. 
The \textit{BJ-RF+lgtc} and \textit{DR-RF+lgtc} also share the same tree structures (see  Figures 
\ref{fig-of-outcen-bj-lgtc} and \ref{fig-of-outcen-dr-lgtc}) and identify 
nearly the same risk groups as the other censored data methods, 
the difference being that the secondary split on age occurs at 70.5 rather than 71.5. 
Importantly, the trees for out-field failure respectively built using the uncensored and censored datasets are very similar. 
Figure \ref{fig4} respectively summarize the estimated CIFs obtained using the uncensored, \textit{DR-RF(npar)},
\textit{IPCW}$_{\!2}$ and \textit{BJ-RF(npar)} loss functions. There are slight differences in the latter three estimators
due to the way in which the CIFs (i.e., terminal nodes) are estimated. We see that the three highest risk groups are 
identified as being the same for all methods; the differences created by
censoring essentially occur in the risk determination 
for the oldest squamous patients, where censoring tends to be the heaviest. 
%
%
In general, the results show that younger patients with non-squamous histology have the highest risk of out-field failure. This observation is
important,  as these patients may be considered as candidates for more intensified treatment regimens.

\begin{figure}[!ht]
\caption{
Risk stratification for death without progression. 
Panels contain estimated CIFs obtained using the following composite loss
functions: \textit{Uncensored} [top left]; \textit{DR-RF(npar)} [top right]; \textit{IPCW$_2$} [bottom left]; 
\textit{BJ-RF(npar)} [bottom right]. CIF estimates are calculated
at the following percentiles of observed all-cause event time:
15, 25, 35, 50, 65, 75, 80. Estimates are otherwise interpolated.}
\centering
\includegraphics[scale=0.9]{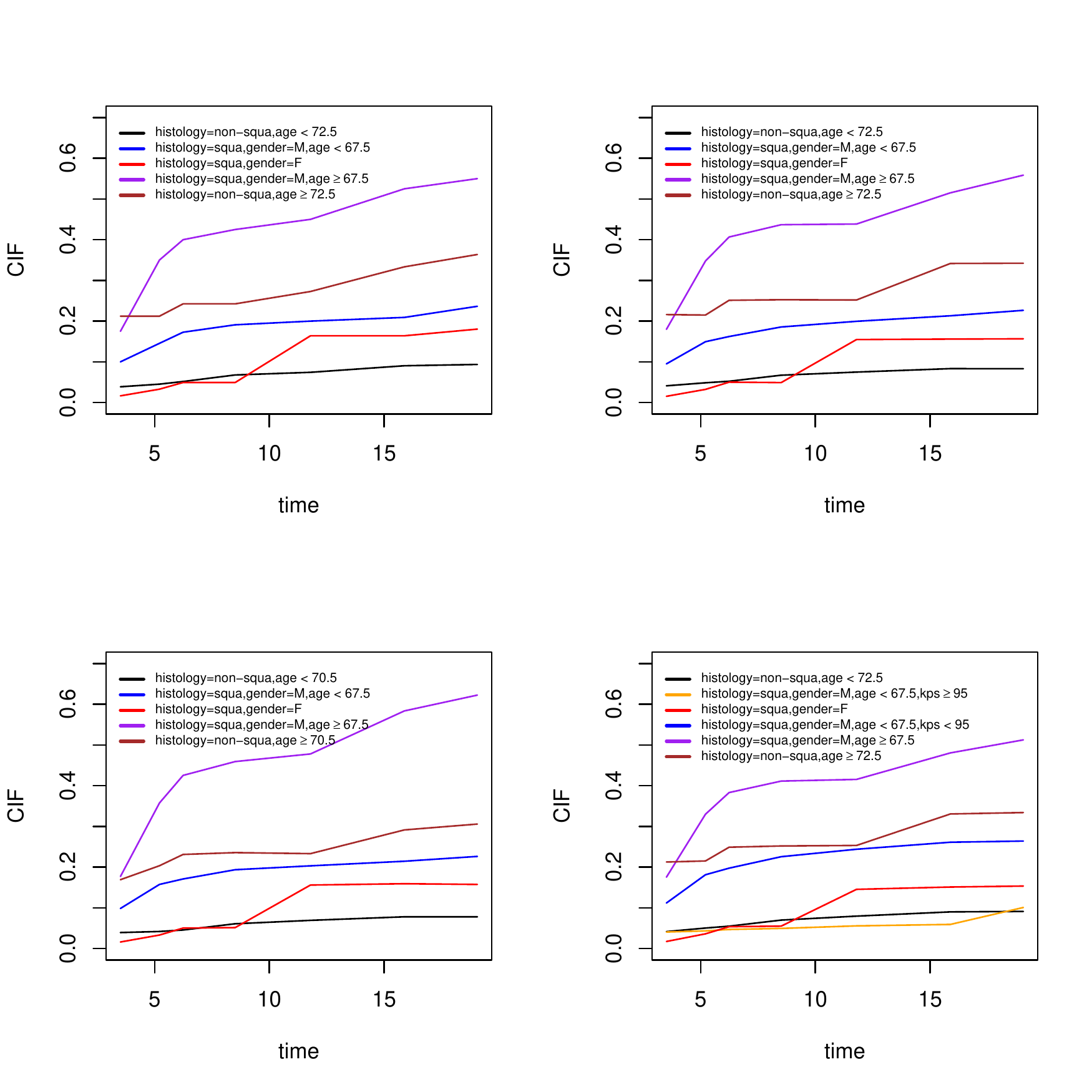}
\label{fig5}
\end{figure}

For death without progression, the fitted tree from uncensored data creates 5 risk groups: non-squamous histology with age $<$ 72.5; male, with age $<67.5$ and squamous histology; male, with age $\geq$ 67.5 and squamous histology; female with squamous histology; and, non-squamous histology with age $\geq$ 72.5.  This tree is given in Figure \ref{fig-death-u} in the Appendix. The tree built using \textit{IPCW}$_{\!1}$ results in a root node. The trees built with \textit{IPCW}$_{\!2}$ and the two doubly robust methods are nearly identical to that built using the uncensored dataset, the difference being that the age cutpoint of 72.5 is replaced by 70.5; see Figures \ref{fig-death-cen-ipcw2} $-$ \ref{fig-death-cen-dr-rf-lgtc} of the Appendix.  The tree built using \textit{BJ-RF(npar)} is also very similar, but incorporates an extra cutpoint on Karnofsky score for younger males with squamous histology; this distinguishes those having a score of 100 from those having a lower score (see Figure \ref{fig-death-cen-bj-rf-npar} of the Appendix).    The tree built using the \textit{BJ-RF+lgtc} method differs somewhat more substantially for younger males with squamous histology, splitting on both Karnofsky score (70 or 80 vs.\ 90 or 100) \and treatment (RX = 1 vs.\ not); see Figure \ref{fig-death-cen-bj-rf-lgtc} of the Appendix.   The comparison of these trees points to a benefit of using the doubly robust loss over the Buckley-James loss, as the censoring distribution is modeled correctly; the comparison of the augmented loss results to those obtained for \textit{IPCW}$_{\!1}$ (where censoring is modeled correctly) further highlights the value of incorporating additional information into the model building process. Figure \ref{fig5} is obtained analogously to 
Figure \ref{fig4} and shows the estimated CIFs and risk stratification
obtained using the uncensored and censored datasets. For death without progression, the risk evidently increases 
with increasing age and being male; the risk also increases with squamous histology, though in a way that appears to be age-dependent.

\section{Discussion}
\label{discuss}

Trees remain of significant interest to practitioners, especially so clinicians.  The proposed doubly robust CIF regression tree demonstrates improved performance compared to IPCW-based methods whether a single time point or composite loss function is used, with composite loss functions giving much better performance overall.  To our knowledge, there are no publically available software packages that directly implement tree-based regression methods for competing risks; our proposed methods can be implemented using existing software, with example code made available as part of supplementary materials. 

Several extensions are possible. For example,  it is easy in principle to combine the proposed estimation procedure with ensemble methods, providing an interesting alternative to the random forests methods recently introduced by \cite{mogensen2013random}  and \cite{ishwaran2014random}.  With no random feature selection, this extension is easily  accomplished  through bagging (i.e., bootstrap-based aggregation). However, with random feature selection,  new software is required, as existing software does not provide for the same possibility of extending {\sf rpart} for use with other loss functions. A second interesting direction is to extend both regression tree and ensemble procedures to the problem of simultaneous estimation of multiple CIFs. Similarly to the case of a single CIF, we anticipate that such multivariate problems can be handled using existing software in the case of regression trees 
and that new software will again be needed for related ensemble extensions.
 
\section*{Acknowledgments}
This work was supported by the National Institutes of Health (R01CA163687: AMM, RLS, YC; U10-CA180822: CH). 
We thank the NRG Oncology Statistics and Data Management Center for providing de-identified RTOG 9410 clinical trial data
under a data use agreement. 



\clearpage


\newpage

\appendix

\singlespace
\begin{center}
{\Large \bf
Supplementary Material for \\
{\em Regression Trees for Cumulative Incidence Functions} \\[1ex]
by Cho, Molinaro, Hu and Strawderman \\[2ex]
}
\end{center}
\section{Appendix}

\numberwithin{equation}{section}
\setcounter{equation}{0} \renewcommand{\theequation}{A.\arabic{equation}} 

\setcounter{table}{0}
\renewcommand{\thetable}{A.\arabic{table}}

\setcounter{figure}{0}
\renewcommand{\thefigure}{A.\arabic{figure}}

\setcounter{page}{1}
\renewcommand{\thepage}{(A-\arabic{page})}

References to figures, tables, theorems and equations preceded by ``A.'' are internal to this appendix; all
other references refer to the main paper.
		
\subsection{Equivalence of augmentation for \textit{IPCW}$_{\! 1}$ and \textit{IPCW}$_{\! 2}$}

\label{loss-equiv-dr}
\noindent The equivalence result to be proved below holds for any pair of suitable models of $G$ and $\Psi$, 
including $G_0$ and $\Psi_0$. 
Throughout, we suppose that suitable models $\Psi$ and $G(\cdot|\cdot)$ are given. The proof will be facilitated
by first establishing that \textit{IPCW}$_{\! 2}$ can be rewritten as an IPCW estimator in standard form and
augmented similarly to \textit{IPCW}$_{\! 1}.$ We will then show that the difference between its
augmented version and that for \textit{IPCW}$_{\! 1}$ are mathematically identical.

Define the modified data ${\cal O}(t) = (\tilde T_i(t), \Delta_i(t), W_i),i=1,\ldots,n,$
where
$\Delta_i(t) = I(T_i(t) \leq C)$ and $T_i(t) = T_i \wedge t$ for $i=1,\ldots,n.$ 
We first observe that the \textit{IPCW}$_{\! 2}$ loss 
can be re-expressed as a standard IPCW estimator that can be computed from
${\cal O}(t).$ In particular, the \textit{IPCW}$_{\! 2}$ loss (i.e.,
\eqref{eq:2} with $t^*=t$)
is mathematically equivalent to
\begin{gather}
\label{ipcw2 rewrite}
L_{m,t}^{ipcw}({\cal O}(t),\psi_m; G)  = 
\frac{1}{n}\sum_{i=1}^{n}\sum_{l=1}^L I\{W_i \in \mathcal{N}_l\}\bigg{[}\frac{\Delta_i(t)\{
I( \tilde T_i(t)< t, M_i=m) - \beta_{lm}(t)\}^2}{G(\tilde T_i(t)|W_i)}\bigg{]}.
\end{gather}
This equivalence follows because 
$I( \tilde T_i \leq t, M_i=m) = I( \tilde T_i(t) < t, M_i=m),  i=1,\ldots,n$ almost surely provided that
$\tilde T_i$ is continuous. Strict inequality on the right-hand side is necessary here because
the fact that $\tilde T_i(t) \leq t$ almost surely implies $0 = I( \tilde T_i \leq t, M_i=m) \neq I( \tilde T_i(t) \leq t, M_i=m) = 1$  
whenever $\tilde T_i > t$ and $M_i = m.$ 
The critical observation is that the loss estimator \eqref{ipcw2 rewrite} is now an IPCW estimator in standard form 
constructed from ${\cal O}(t),$ and thus can be augmented similarly to \textit{IPCW}$_{\! 1}$
\citep[cf.\ ][Sec.\ 9.3, 10.3]{tsiatis2007semiparametric}.
In particular, the augmented loss is given by
$L_{m,t}^{dr}({\cal O}(t),\psi_m;  G_0,\Psi_0) = L_{m,t}^{ipcw}({\cal O}(t),\psi_m;  G_0) 
+ L_{m,t}^{aug}({\cal O}(t),\psi_m; G_0, \Psi_0),$ 
where 
\[
L_{m,t}^{aug}(\mathcal{O}(t),\psi_m;  G, \Psi) =
 \frac{1}{n}\sum_{l=1}^{L}\sum_{i=1}^{n}I\{W_i \in \mathcal{N}_l\}
\int_0^{\tilde{T}_i(t)} \frac{V_{lm}(u;t,W_i,\Psi)}{G(u|W_i)}dM_G(u|W_i).
\]
Note that this last expression is just \eqref{eq:4a1}, but with
an upper limit of integration of $\tilde{T}_i(t)$ in place of $\tilde T_i$.

We can now establish the equivalence of the augmented loss functions respectively derived from
 the \textit{IPCW}$_{\! 1}$ and \textit{IPCW}$_{\! 2}$ losses. Trivially, we may rewrite
\begin{gather}
\label{ipcw1}
L_{m,t}^{ipcw}({\cal O},\psi_m; G) = \frac{1}{n} \sum_{i=1}^n \sum_{l=1}^L I\{W_i \in \mathcal{N}_l\}
\bigg{[}\frac{\Delta_i\{Z_{im}(t) - \beta_{lm}(t)\}^2}{G(T_i|W_i)}   \bigg{]}
\end{gather}
and 
\begin{gather}
\label{ipcw2}
L_{m,t}^{ipcw}({\cal O}(t),\psi_m; G)  = 
\frac{1}{n}\sum_{i=1}^{n}\sum_{l=1}^L I\{W_i \in \mathcal{N}_l\}\bigg{[}\frac{\Delta_i(t)\{
Z_{im}(t) - \beta_{lm}(t)\}^2}{G(T_i(t)|W_i)}\bigg{]}.
\end{gather}

\noindent Consequently, $L_{m,t}^{dr}({\cal O},\psi_m;  G,\Psi) - L_{m,t}^{dr}({\cal O}(t),\psi_m;  G,\Psi) 
= (A) - (B)$ almost surely, where
\begin{gather*}
(A) =  \frac{1}{n}\sum_{l=1}^{L}\sum_{i=1}^{n}I\{W_i \in \mathcal{N}_l\}\bigg{(}\frac{\Delta_i}{G(T_i|W_i)} - \frac{\Delta_i(t)}{G(T_i(t)|W_i)}\bigg{)}\{Z_{im}(t) - \beta_{lm}(t)\}^2 
\end{gather*}
and
\begin{gather*}
(B) =  \frac{1}{n}\sum_{l=1}^{L}\sum_{i=1}^{n}I\{W_i \in \mathcal{N}_l\} 
 \int_{\tilde{T}_i(t)}^{\tilde{T}_i} \frac{V_{lm}(u;t,W_i,\Psi)}{G(u|W_i)}dM_G(u|W_i).
\end{gather*}
\noindent Consider the integral term appearing in (B) for any fixed value of $i$. By the definition of $\tilde{T}_i(t)$, the integral is clearly zero if $\tilde{T}_i(t) = \tilde{T}_i$ (i.e., when $\tilde T_i < t$); hence, we only need to consider the calculation of
\[
 \int_{t}^{\tilde{T}_i} \frac{V_{lm}(u;t,W_i,\Psi)}{G(u|W_i)}dM_G(u|W_i).
 \]
 for $\tilde T_i \geq t.$ Recall that $V_{lm}(u;t,W_i,\Psi) = E_{\Psi}\{Z_{im}(t) - \beta_{lm}(t) | T \geq u, W_i\}^2
  = y_m(u;t,w,\Psi) - 2y_m(u;t,w,\Psi)\beta_{lm}(t) + \beta_{lm}^2(t)$, where $y_m(u;t,w,\Psi)$ is given in 
  \eqref{eq:2b}. By definition, we have that $y_m(u;t,w,\Psi) = 0$ for $u \geq t$; hence, for $\tilde T_i \geq t,$
\begin{gather}
\label{eq:1c}
\int_{t}^{\tilde{T}_i} \frac{V_{lm}(u;t,W_i,\Psi)}{G(u|W_i)}dM_G(u|W_i) = \beta_{lm}^2(t) \int_{t}^{\tilde{T}_i} \frac{dM_G(u|W_i)}{G(u|W_i)}.
\end{gather}
Calculations analogous to those done in \cite{bai2013doubly} show that
\begin{gather}
\label{eq:1d} 
\int_t^{\tilde{T}_i} \frac{dM_G(u|W_i)}{G(u|W_i)} = 
 I(\tilde{T}_i \geq t) \bigg{(}\frac{1}{G(t|W_i)} - \frac{\Delta}{G(\tilde{T}_i|W_i)}  \bigg{)}.
\end{gather}
Substituting (\ref{eq:1d}) into (\ref{eq:1c}) gives
\begin{gather}
\label{eq:1e}
\int_{t}^{\tilde{T}_i} \frac{V_{lm}(u;t,W_i,\Psi)}{G(u|W_i)}dM_G(u|W_i) = \beta_{lm}^2(t) I(\tilde{T}_i \geq t) \bigg{(}\frac{1}{G(t|W_i)} - \frac{\Delta_i}{G(\tilde{T}_i|W_i)}\bigg{)}.
\end{gather}
Hence we can see that 
\[
L_{m,t}^{dr}({\cal O},\psi_m;  G,\Psi) -  L_{m,t}^{dr}({\cal O}(t),\psi_m;  G,\Psi) 
= \frac{1}{n}\sum_{i=1}^n I(W_i \in \mathcal{N}_l) \, \Omega_i,
\]
almost surely, where
\[
\Omega_i  = \bigg{(}\frac{\Delta_i}{G(T_i|W_i)} - \frac{\Delta_i(t)}{G(T_i(t)|W_i)}\bigg{)}\{Z_{im}(t) - \beta_{lm}(t)\}^2 
+ I(\tilde{T}_i \geq t) \bigg{(}\frac{1}{G(t|W_i)} - \frac{\Delta_i}{G(\tilde{T}_i|W_i)} \bigg{)} \beta_{lm}^2(t).
\]

\noindent Recall that $T_i$ and $C_i$ are both continuous; hence, 
$P(T_i = t) = P(C_i = t) = 0$. We now consider the calculation of $\Omega_i$
under the 6 possible ways in which $T_i,$ $C_i$ and $t$ can
be ordered with positive probability:
\begin{enumerate}
\item \underline{$T_i < C_i < t:$}
In this case,  $\Delta_i = \Delta_i(t) = 1,$ $T_i(t) = T_i,$ and $\tilde{T}_i < t.$
Hence,
\[
 \Omega_i  = \bigg{(}\frac{1}{G(T_i|W_i)} - \frac{1}{G(T_i|W_i)}\bigg{)}\{Z_{im}(t) - \beta_{lm}(t)\}^2 
+ 0 \cdot \bigg{(}\frac{1}{G(t|W_i)} - \frac{1}{G(\tilde{T}_i|W_i)} \bigg{)} \beta_{lm}^2(t) = 0.
\]

\item \underline{$T_i < t < C_i:$}
As in Case 1, $\Delta_i = \Delta_i(t) = 1,$ $T_i(t) = T_i,$ and $\tilde{T}_i < t;$
hence, $\Omega_i = 0.$
\item \underline{$t < T_i <  C_i:$} In this case, $\Delta_i = 1,$ $T_i(t)=t,$ $\Delta_i(t)= I(T_i(t) \leq C_i)
= I(t \leq C_i) = 1$ and $\tilde T_i >t$. 
Moreover, $Z_{im}(t)=I(T_i \leq t, M_i = 1) = 0$ regardless of $M_i$ because $T_i > t.$ Hence,
\begin{gather*}
\Omega_i = 
\bigg{(} \frac{1}{G(T_i|W_i)} - \frac{1}{G(t|W_i)} \bigg{)} \{0-\beta_{lm}(t)\}^2 
+ \beta_{lm}^2(t) \bigg{(}\frac{1}{G(t|W_i)} - \frac{1}{G(T_i|W_i)}\bigg{)} \\
= \frac{\beta_{lm}^2(t)}{G(T_i|W_i)} - \frac{\beta_{lm}^2(t)}{G(t|W_i)} + \frac{\beta_{lm}^2(t)}{G(t|W_i)} - \frac{\beta_{lm}^2(t)}{G(T_i|W_i)} = 0.
\end{gather*}
\item \underline{$C_i < T_i < t:$} In this case, $\Delta_i= 0,$ $\tilde T_i < t,$
 $T_i(t) = T_i,$ and  $\Delta_i(t)= I(T_i(t) \leq C_i) = 0.$ Hence,
 \[
 \Omega_i  = \bigg{(}\frac{0}{G(T_i|W_i)} - \frac{0}{G(T_i|W_i)}\bigg{)}\{Z_{im}(t) - \beta_{lm}(t)\}^2 
+ 0 \times \bigg{(}\frac{1}{G(t|W_i)} - \frac{0}{G(\tilde{T}_i|W_i)} \bigg{)} \beta_{lm}^2(t) = 0.
\]
\item 
\underline{$C_i < t < T_i:$ }
In this case, $\Delta_i = 0,$ $\tilde T_i < t,$ $\Delta_i(t)= I(T_i(t) \leq C_i) = 0$ and $T_i(t) = t.$
Hence, 
\[
 \Omega_i  = \bigg{(}\frac{0}{G(T_i|W_i)} - \frac{0}{G(t|W_i)}\bigg{)}\{Z_{im}(t) - \beta_{lm}(t)\}^2 
+ 0 \times \bigg{(}\frac{1}{G(t|W_i)} - \frac{0}{G(\tilde{T}_i|W_i)} \bigg{)} \beta_{lm}^2(t) = 0.
\]
\item 
\underline{$t < C_i < T_i:$} 
In this case, $\Delta_i = 0,$ $\tilde T_i > t,$ $\Delta_i(t)=1,$ $T_i(t) = t$ and $Z_{im}(t) = 0$
regardless of $M_i$ because $T_i > t$. Hence,
\begin{gather*}
\Omega_i = 
\bigg{(}\frac{0}{G(T_i|W_i)} - \frac{1}{G(t|W_i)} \bigg{)} \{0-\beta_{lm}(t)\}^2 
+ \beta_{lm}^2(t) \bigg{(}\frac{1}{G(t|W_i)} - \frac{0}{G(T_i|W_i)}\bigg{)} \\
= 
- \frac{\beta_{lm}^2(t)}{G(t|W_i)} + \frac{\beta_{lm}^2(t)}{G(t|W_i)} = 0.
\end{gather*}
\end{enumerate}
In summary, we have proved that
$L_{m,t}^{dr}({\cal O},\psi_m;  G,\Psi) - L_{m,t}^{dr}({\cal O}(t),\psi_m;  G,\Psi) = 0$
except possibly on a set which has probability measure 0.

\subsection{Further details on CART algorithms using augmented loss}
\label{dimplement}
The extension of the CIF regression tree methodology 
described in Section \ref{implement0} to the case where ${\cal O}$ is observed has two main steps:
\begin{enumerate}
\item Throughout, replace the $L_2$ loss function used by CART with  $L_{m,t}^{dr}({\cal O},\psi_m; \hat G, \hat \Psi);$
\item Use the modified algorithm in Step \#1 to  grow a maximal tree, thereby generating a sequence of trees as candidates for the best tree;
\item Using cross-validation, select the best tree from this sequence via cost complexity pruning.
\end{enumerate}
Regarding Step \#1, the overall structure of the CART algorithm is independent of the choice of loss function; the specification of the loss
function only plays a role in determining how the splitting, evaluation, and model selection decisions are made in Steps \#2 and \#3 \citep[cf.,][]{breiman1984classification}.

Step \#2 serves to restrict the search space, that is, the set of possible trees that one must consider in determining
which tree is optimal; see below.
To describe how Step \#3 is carried out, we first need to define the notion of a cross-validated risk estimator.
Let ${\cal O} = \cup_{q=1}^{Q} {\cal O}_q$ be a partition
of ${\cal O}$ such that each ${\cal O}_q$ contains all observed data on some subset of subjects and each
subject in ${\cal O}$ appears in exactly one of the sets ${\cal O}_1,\ldots,{\cal O}_Q$.
Let $p_q = \frac{1}{n}\sum_{i=1}^{n} S_{i,q},$ where 
$S_{i,q} = 1$ if observation $i$ is in dataset ${\cal O}_q$ and zero otherwise. 
Recalling \eqref{simple-V}, define the modified notation
\[
\hat V_{lm}(u;t,w,\psi_{m}) = y_{m}(u;t,w, \hat \Psi) - 2 y_{m}(u;t,w, \hat \Psi) \psi_{m}(t; w) + [\psi_{m}(t; w)]^2;
\]
as given, $\hat V_{lm}(u;t,w,\psi_{m})$ is now a function of $\psi_{m}(t; w)$.
Fixing $q$ and for each $i \in {\cal O}_q,$ let
\[
\varphi_{i,m}^{dr}(\mathcal{O}_q,\hat{\psi}^{(q)}_m(t;W_i)) = \frac{\Delta_i\{\tilde{Z}_{im}(t) - \hat{\psi}^{(q)}_m(t;W_i)\}^2}{\hat{G}(\tilde{T}_i|W_i)} + \displaystyle \int_{0}^{\tilde{T}_i} \frac{\hat V_{lm}(u;t,W_i,\hat{\psi}^{(q)}_m)}{\hat{G}(u|W_i)} dM_{\hat G}(u|W_i)
\]
where $\hat{\psi}^{(q)}_m(t;w)$ is the prediction obtained from a tree that estimates $\psi_{0m}(t;w)$ 
from the reduced dataset ${\cal O} \setminus {\cal O}_q.$ Then, similarly to
\citet{steingrimsson2016doubly}, a cross-validated doubly 
robust risk estimator can be defined as follows:
\[
\hat {\cal R}(t;\hat{\psi}^{(1)}_m, \ldots, \hat{\psi}^{(Q)}_m) =  \frac{1}{nQ} \sum_{q=1}^{Q} 
				\sum_{i=1}^{n} \frac{I(S_{i,q}=1)}{p_q}
~\varphi_{i,m}^{dr}(\mathcal{O}_q,\hat{\psi}^{(q)}_m(t;W_i)). 
\]

Returning to Step \#3: Suppose Steps \#1 and 2 have been run. The corresponding unpruned maximal
tree generates a sequence of (say) $R$ subtrees, each of which is a candidate for the final tree. Each of
these trees can be identified from this maximal tree by a unique choice of the so-called cost complexity
tuning penalty parameter; call these parameter values $\alpha_1,\ldots,\alpha_R$.
Now, for each fixed $r = 1,\ldots, R,$ let $\hat{\psi}^{(1,r)}_m, \ldots, \hat{\psi}^{(Q,r)}_m$ denote the sequence of trees
obtained by running the doubly robust tree building procedure on the datasets ${\cal O} \setminus {\cal O}_q, q =1,\ldots,Q,$ where 
each such tree is determined by cost complexity pruning using the penalty parameter $\alpha_r$. Define
$\hat {\cal R}(t;\hat{\psi}^{(1,r)}_m, \ldots, \hat{\psi}^{(Q,r)}_m)$ to be the corresponding cross-validated 
risk estimates. 
The final tree is then obtained from the initial maximal tree using the penalty parameter $\alpha_{\hat r},$ where $\hat{r}$ 
minimizes $\hat {\cal R}(t;\hat{\psi}^{(1,r)}_m, \ldots, \hat{\psi}^{(Q,r)}_m), r= 1,\ldots,R.$

The indicated procedure is modified in an obvious way to accommodate IPCW-type and Buckley-James Brier loss functions.
In the case of the corresponding composite loss function, a weighted sum of the desired loss estimates over $t_1,\ldots,t_J$ is instead used.  Beyond these loss modifications, estimation of the maximal tree and selection of the corresponding best tree is carried out exactly as described above. In practice, implementation of the algorithm just described is possible using {\tt rpart}'s capabilities for incorporating user-written splitting and evaluation functions \citep{therneau2015introduction}. Examples of such code are provided as part of the Supplementary Materials for this paper.

\subsection{Estimating $\Psi$ for computing augmented loss functions}
\label{sec:est-CE}

Section \ref{nuisance} discusses methods for estimating both the censoring distribution $G$ and the parameter $\Psi$
when computing observed data loss functions.  This section expands on the four methods for estimating $\Psi,$
needed for computing \eqref{eq:2b}, that are described in Section \ref{nuisance}. We assume $K=2.$

\begin{enumerate}
\item \textit{RF(npar)}: for $m=1,2,$ $\psi_{0m}(\cdot|\cdot)$ is estimated using 
nonparametric random forests. More specifically, the {\tt rfsrc} function in the {\sf randomForestSRC} package 
is used \citep{ishwaran2014random}, where the \verb|predict| command is used to estimate 
$\psi_{01}(u|w)$ and $\bar \psi_0(u|w)$ for $K=2$ and $u \geq 0$. For $m=1,2$ and $B=500$ bootstrap samples,
the components of $\hat \Psi$ are given by 
$\hat{\psi}_m^{\mbox{\tiny rf}}(u|w) = B^{-1} \sum_{b=1}^{B} \hat{\psi}_{m[b]}(u|w);$
here, for the $b^{th}$ bootstrap sample, $\hat{\psi}_{1[b]}(u|w) = \hat{P}_{[b]}(T \leq u, M=1 | W=w)$ is the estimated CIF of interest and 
$\hat{\psi}_{2[b]}(u|w) = \hat{P}_{[b]}(T \leq u, M \neq 1 | W=w)$ is the corresponding "all other causes" estimate.
The event-free survival function is estimated in the obvious manner.

\item \textit{RF+lgtc}: for $m=1,2,$ $\psi_{0m}(\cdot|\cdot)$ is estimated using ``random parametric forest''.
That is, as above, the {\tt rfsrc} function in the {\sf randomForestSRC} package is 
first used \citep{ishwaran2014random} to construct a forest (i.e., $B=500$ trees) with $K=2$. 
Using the tree obtained for the $b^{th}$ bootstrapped sample,  
$\hat{\psi}_{m[b]}(u|w) = \hat{P}_{[b]}(T \leq u, M=m | W=w), m=1,2$ 
are then estimated by fitting the three parameter logistic cumulative
incidence model of \cite{cheng2009modeling} separately to the data falling into each terminal 
node. Finally, similarly to $\hat{\psi}_m^{\mbox{\tiny rf}}(u|w),$ the desired ensemble
predictors are obtained by computing bootstrap averages.

\item \textit{FG(true)}: the simulation model 
of Section \ref{sim set} is an example of the class of models considered in 
\citet{jeong2007parametric}. For $m=1,2,$ $\psi_{0m}(\cdot|\cdot)$ are estimated
by maximum likelihood under a correctly specified parametric model. 

\item \textit{lgtc+godds}: for $m=1,2,$ $\psi_{0m}(\cdot|\cdot)$ is estimated by maximum likelihood
using the parametric cumulative incidence model proposed in \cite{shi2013constrained}.
\end{enumerate}

\subsubsection{The CIF models of \cite{cheng2009modeling} \cite{shi2013constrained}}
\label{three par details}

Here, we describe the parametric models of \cite{cheng2009modeling} \cite{shi2013constrained} that are
used by \textit{RF+lgtc} and \textit{lgtc+godds}. Specifically, assuming $K=2$ and hence for $m=1,2,$ 
these models are given below:
\begin{enumerate}
\item Three parameter logistic model without covariates \citep[\textit{RF+lgtc};][]{cheng2009modeling}: \\[1.5ex]
This model is motivated by nonlinear modeling in bioassay and dose-response curves \citep{ritz2005bioassay}.
Specifically, the parametric model that is fit by maximum likelihood separately within each terminal node of a tree is given by 
\begin{gather*}
\psi_m(t;\xi_m) = \frac{p_m\exp\{b_m(t-c_m)\} - p_m\exp(-b_m c_m)}{1+\exp\{b_m(t-c_m)\}} 
\end{gather*}
where $\xi_m = (b_m,c_m,p_m)^T$ consists of three parameters: $p_m$ is upper asymptote of $\psi_m(\cdot;\cdot)$ when $t \rightarrow \infty$; and, $b_m$ and $c_m$ are the slope and center of the curve \citep{cheng2009modeling}. To satisfy the additivity constraint required of CIFs, 
we further assume $p_2=1-p_1$. 
\item Three parameter logistic model with covariates \citep[\textit{lgtc+godds};][]{shi2013constrained}: \\[1.5ex]
\cite{shi2013constrained} extends the model of \cite{cheng2009modeling} to incorporate
covariates. Specifically, the parametric model used for $\psi_{0m}(t;w)$ is 
\begin{gather*}
\psi_m(t;\mathbf{W},\boldsymbol{\gamma},\xi_m) = g_m^{-1}[g_m\{\psi_{m}(t;\xi_m)\} + \boldsymbol{\gamma}^T\mathbf{W}]
\end{gather*}
where $g_m$ is a specified nondecreasing function and $\psi_{m}(t;\xi_m)$ is the covariate-independent
model of \cite{cheng2009modeling}. One possible form of $g_m$ is
\begin{gather*}
g_m(u;\alpha_m) = \log[\{(1-u)^{-\alpha_m} - 1\}/\alpha_m], 0 < \alpha_m < \infty
\end{gather*} 
where $\alpha_m$ is parameter for each event type. This model is an extension of the generalized odds-rate model \citep{dabrowska1988estimation}. When using a generalized odds model as a link function $g_m(\cdot;\cdot)$, the CIF for $m=1$
is given by the formula $\psi_1(t;\boldsymbol{W}) = 1-H(t;\boldsymbol{W})^{-1/\alpha_1},$ where
		\begin{gather*}
		H(t;\boldsymbol{W}) = \bigg{\{}\bigg{(}1-\frac{p_1\exp\{b_1(t-c_1) - p_1\exp(-b_1c_1)\}}{1+\exp\{b_1(t-c_1)\}}\bigg{)}^{-\alpha_1} \!\!\!\!-1\bigg{\}}\exp(\boldsymbol{\beta}_1^T\boldsymbol{W})+1;
		\end{gather*}
the CIF for $m = 2$ is given by
		\begin{gather*}
		\psi_2(t;\boldsymbol{W}) = \frac{p_2(\boldsymbol{W})[\exp\{b_2(t-c_2)\} - \exp\{-b_2c_2\}]}{1+\exp\{b_2(t-c_2)\}},
		\end{gather*}
where $p_2(\boldsymbol{W})=1- \psi_1(\infty;\boldsymbol{W}) = [\{(1-p_1)^{-\alpha_1}-1\}\exp(\boldsymbol{\beta}_1^T\boldsymbol{W}))+1]^{-1/\alpha_1}$. Observe that
 the additive constraint $\psi_1(\infty;\boldsymbol{W}) + \psi_2(\infty;\boldsymbol{W}) = 1$ is satisfied. 
 In addition, note that a separate covariate effect for $m=2$  is not modeled and hence there are no additional regression parameters to be estimated (i.e., even though CIF for $m=2$ does depend on covariates through $p_2(\boldsymbol{W})$).   
\end{enumerate} 

\subsection{Miscellaneous Algorithm Specifications}
Two important tuning parameters that govern the size of the maximal tree built in Step \#2 of general CART algorithm are \verb|minbucket|, the minimum possible number of observations in a terminal node, and \verb|minsplit|, the minimum number of observations in a node to be considered for a possible split. Throughout, we set \verb|minbucket|=10 and \verb|minsplit|=30.  For the doubly robust and Buckley-James methods, both uncensored and censored
observations are counted when considering these limits;  for IPCW methods, only uncensored observations are counted.

\subsection{Additional simulation results} 
\label{add sim res}
In this section, we summarize the full set of simulation results using both``single time point" and composite loss functions.
We recall that the IPCW$_1$ and IPCW$_2$ refer to the usual ($t^* = \infty$) and modified ($t^* = t$) IPCW methods, respectively. As
described elsewhere, the Buckley-James (BJ) and doubly robust (DR) methods use several approaches to estimating 
\eqref{eq:2b}, which is required for computing the 
augmented loss function. In particular, as described in Sections \ref{nuisance} (see also 
Section \ref{three par details}), we use nonparametric random forests (RF (npar)), random parametric forests (RF+lgtc), the true data 
generating model  (FG (true)) and the three-parameter logistic distribution in combination with a generalized odds model (Lgtc+godds). 
Results are summarized for each type of loss function. For example, results corresponding to``BJ-RF (npar)" means that we use the 
BJ loss function in combination with nonparametric random forests for calculating \eqref{eq:2b}.

\begin{figure}[ht]
\caption{{Prediction error for event 1 when $n=250$ (multiplied by 100). Losses using a single time point and multiple time points. 
\textit{IPCW}$_{1}$ and \textit{IPCW}$_2$ are time-independent IPCW and time-dependent IPCW, respectively; \textit{BJ-RF(npar)}, \textit{BJ-RF+lgtc}, \textit{BJ-lgtc+godds} and \textit{BJ-FG(true)} are respectively built using Buckley-James loss functions with augmentation term estimated via nonparametric random forests, 
random parametric forests using ensembles of parametric models described in \cite{cheng2009modeling}, the parametric model of \cite{shi2013constrained},
and under the correct (i.e., consistently estimated) simulation model; and, 
\textit{DR-RF(npar)}, \textit{DR-RF+lgtc}, \textit{DR-lgtc+godds} and \textit{DR-FG(true)} are built using the doubly robust loss function, 
respectively using the same methods for calculating the augmentation term as in  \textit{BJ-RF(npar)}, \textit{BJ-RF+lgtc}, \textit{BJ-lgtc+godds} and \textit{BJ-FG(true)}.}}

\begin{minipage}[b][1\textheight][s]{1.1\textwidth}
  \centering
  \includegraphics[height=0.275\textheight,width=\textwidth]{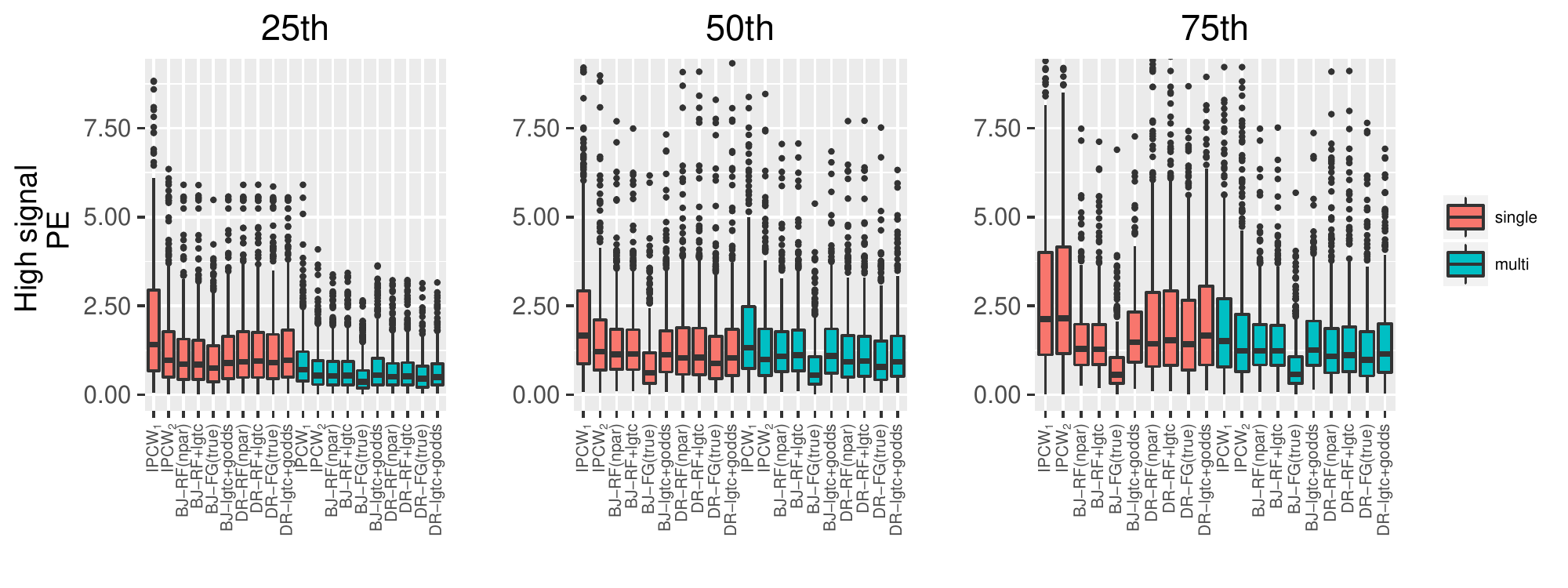}
  \vspace{-0.6cm}
  \includegraphics[height=0.275\textheight,width=\textwidth]{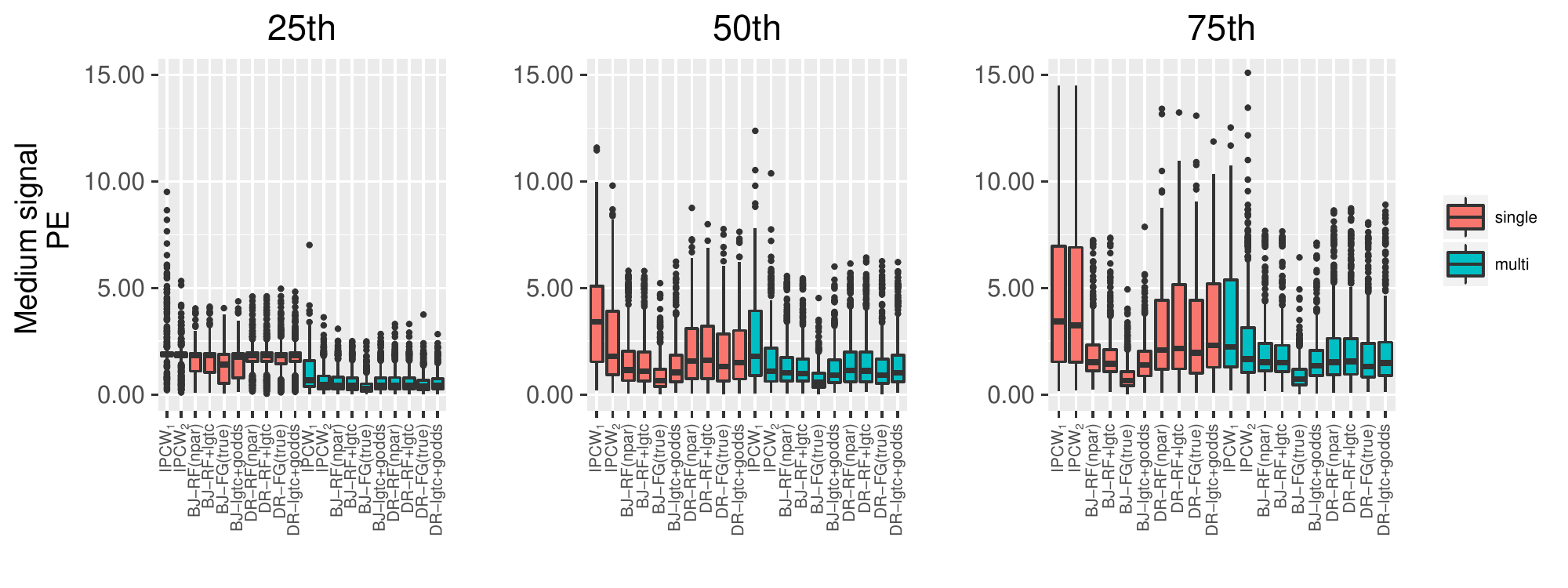}
  \vspace{-0.6cm}
  \includegraphics[height=0.275\textheight,width=\textwidth]{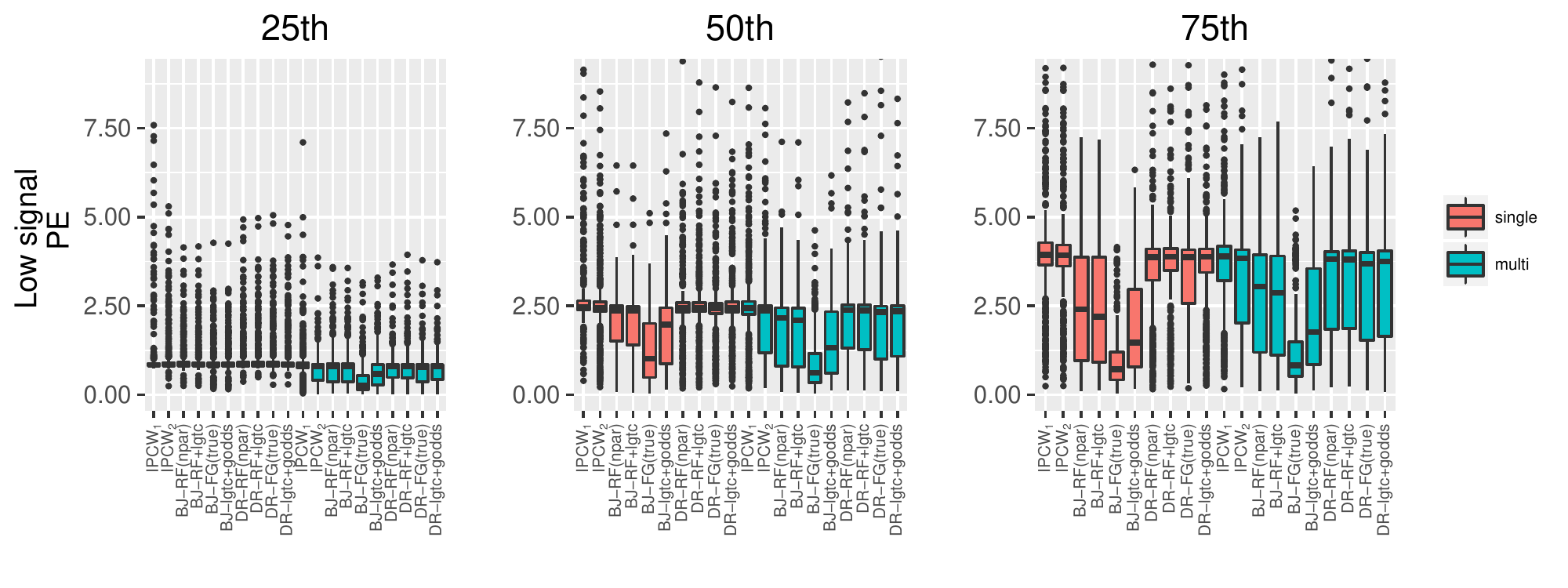}
\end{minipage}
  \vspace{-3cm}
\label{app-fig-1}
\end{figure}

\begin{figure}[ht]
\caption{{Prediction error for event 1 when $n=500$ (multiplied by 100). Losses using a single time point and multiple time points. 
\textit{IPCW}$_{1}$ and \textit{IPCW}$_2$ are time-independent IPCW and time-dependent IPCW, respectively; \textit{BJ-RF(npar)}, \textit{BJ-RF+lgtc}, \textit{BJ-lgtc+godds} and \textit{BJ-FG(true)} are respectively built using Buckley-James loss functions with augmentation term estimated via nonparametric random forests, 
random parametric forests using ensembles of parametric models described in \cite{cheng2009modeling}, the parametric model of \cite{shi2013constrained},
and under the correct (i.e., consistently estimated) simulation model; and, 
\textit{DR-RF(npar)}, \textit{DR-RF+lgtc}, \textit{DR-lgtc+godds} and \textit{DR-FG(true)} are built using the doubly robust loss function, 
respectively using the same methods for calculating the augmentation term as in  \textit{BJ-RF(npar)}, \textit{BJ-RF+lgtc}, \textit{BJ-lgtc+godds} and \textit{BJ-FG(true)}.}}

\begin{minipage}[b][1\textheight][s]{1.1\textwidth}
  \centering
	  \includegraphics[height=0.275\textheight,width=\textwidth]{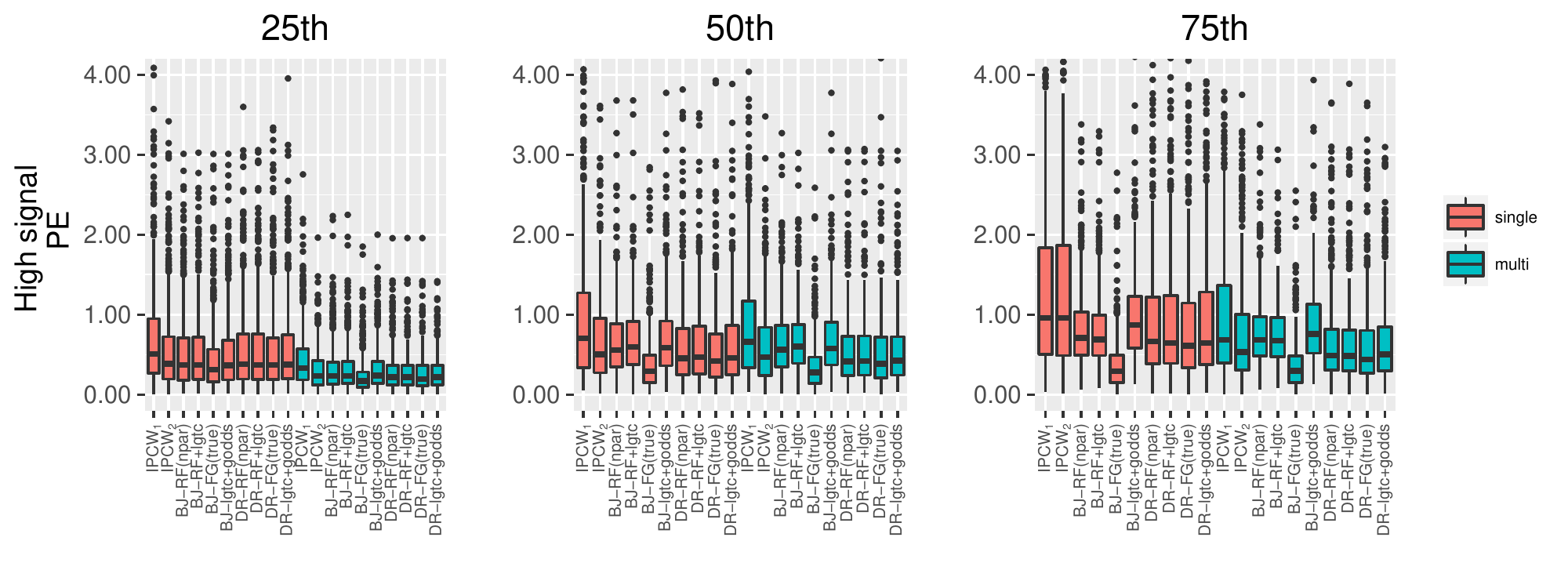}
  \vspace{-0.6cm}
  \includegraphics[height=0.275\textheight,width=\textwidth]{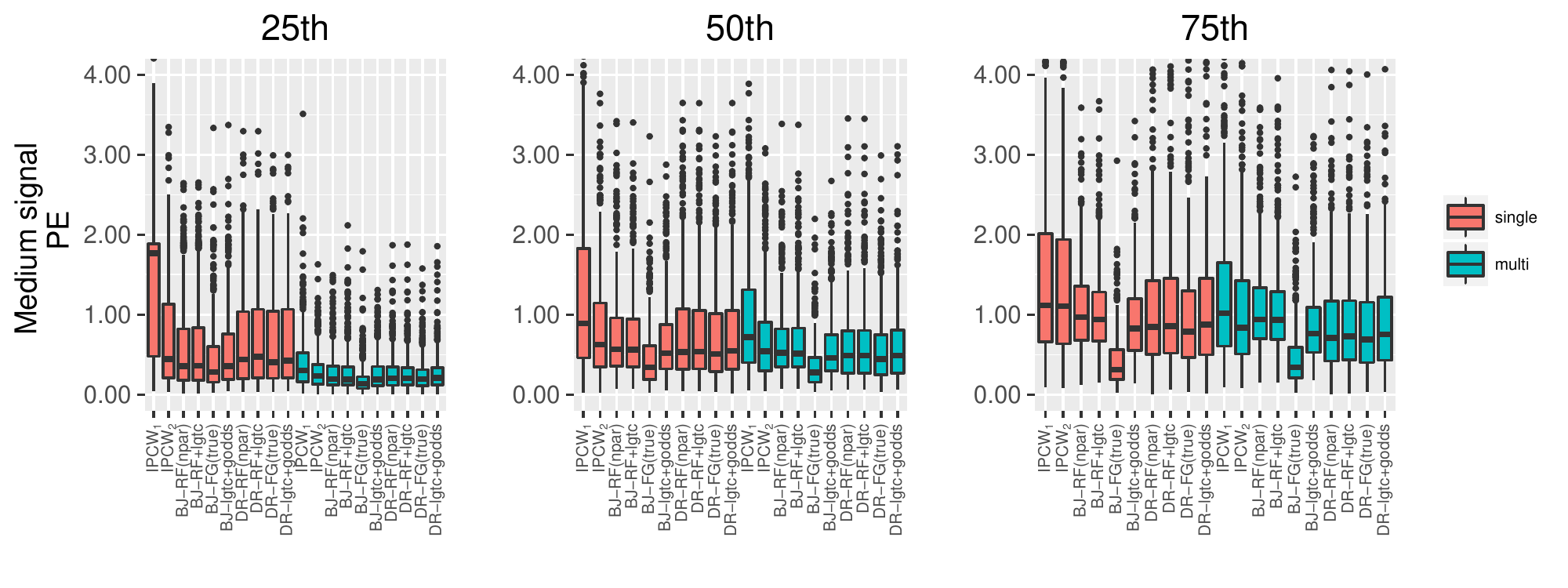}
  \vspace{-0.6cm}
  \includegraphics[height=0.275\textheight,width=\textwidth]{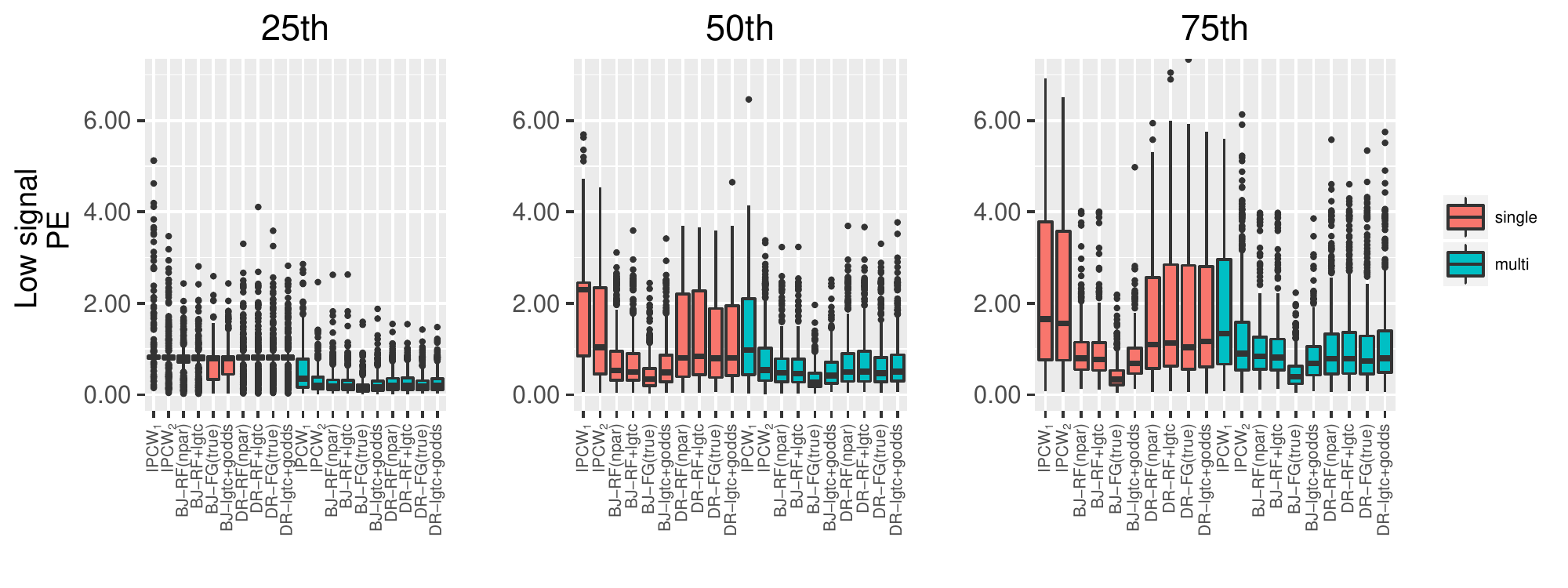}
\end{minipage}
  \vspace{-3cm}
  \label{app-fig-2}
\end{figure}

\begin{figure}[ht]
\caption{{Prediction error for event 1 when $n=1000$ (multiplied by 100). Losses using a single time point and multiple time points. 
\textit{IPCW}$_{1}$ and \textit{IPCW}$_2$ are time-independent IPCW and time-dependent IPCW, respectively; \textit{BJ-RF(npar)}, \textit{BJ-RF+lgtc}, \textit{BJ-lgtc+godds} and \textit{BJ-FG(true)} are respectively built using Buckley-James loss functions with augmentation term estimated via nonparametric random forests, 
random parametric forests using ensembles of parametric models described in \cite{cheng2009modeling}, the parametric model of \cite{shi2013constrained},
and under the correct (i.e., consistently estimated) simulation model; and, 
\textit{DR-RF(npar)}, \textit{DR-RF+lgtc}, \textit{DR-lgtc+godds} and \textit{DR-FG(true)} are built using the doubly robust loss function, 
respectively using the same methods for calculating the augmentation term as in  \textit{BJ-RF(npar)}, \textit{BJ-RF+lgtc}, \textit{BJ-lgtc+godds} and \textit{BJ-FG(true)}.}}

\begin{minipage}[b][1\textheight][s]{1.1\textwidth}
  \centering
  \includegraphics[height=0.275\textheight,width=\textwidth]{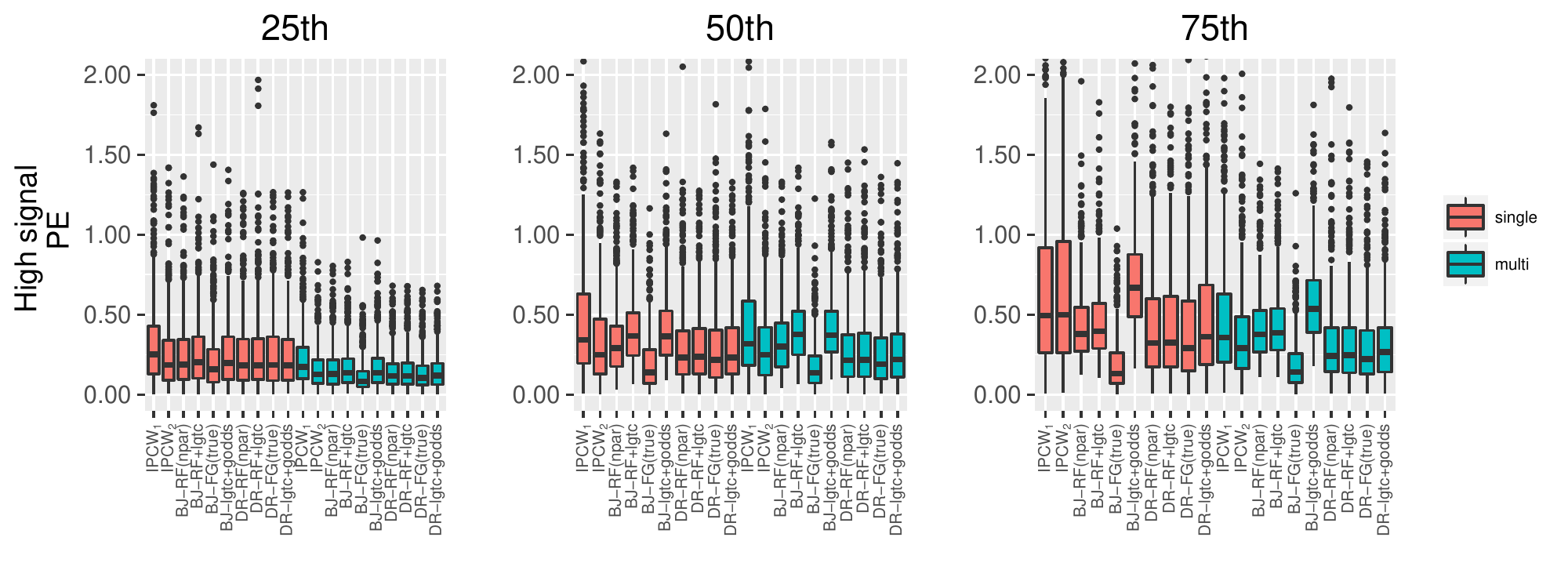}
  \vspace{-0.6cm}
  \includegraphics[height=0.275\textheight,width=\textwidth]{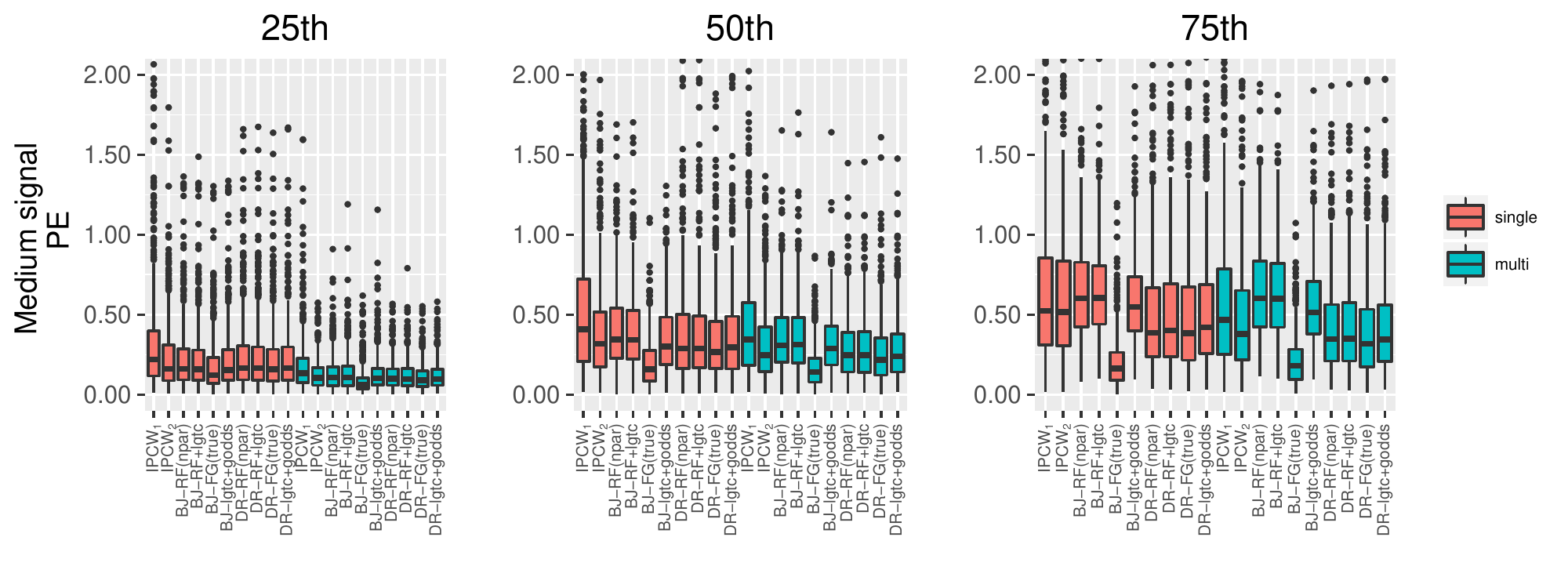}
  \vspace{-0.6cm}
  \includegraphics[height=0.275\textheight,width=\textwidth]{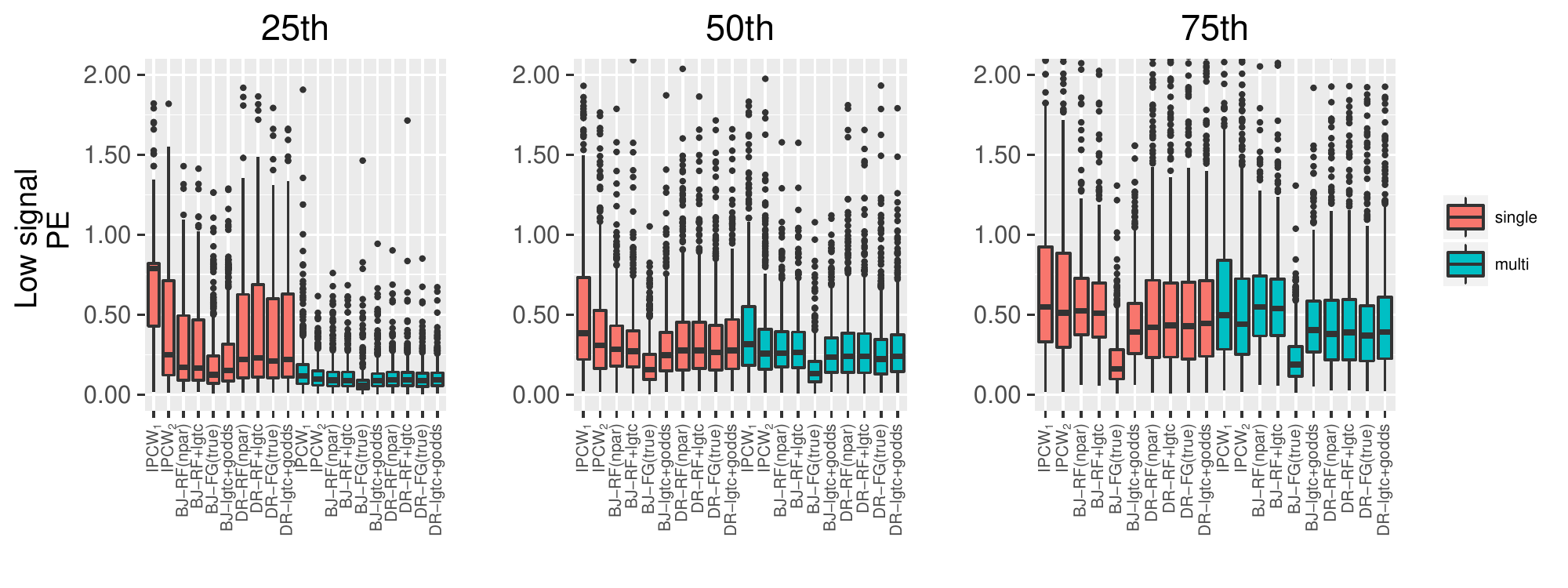}
\end{minipage}
  \vspace{-3cm}
  \label{app-fig-3}
\end{figure}

\indent Table \ref{app-tab-1} -- \ref{app-tab-4} summarize the numerical performance of the fitted trees and Figures \ref{app-fig-1} -- \ref{app-fig-3} 
show boxplots of prediction error.  The choice of time points and method of summarizing the results in the tables and figures is the same as
in Section \ref{main sim res}.
It is easy to see that (i) performance generally improves with increasing sample size and tends to be worst at $t_1$; (ii) there is substantially more variation in results using a single time point loss in comparison to a composite loss; and, (iii) the use of the composite loss function results in better overall performance,
inlcuding lower prediction error, in comparison to the use of any single time point loss function.

In the medium and low signal settings, greater differences between methods emerge, with a clear improvement at $n=500$ compared to $n=250.$ Regardless of sample size, \textit{IPCW}$_1$ has the least favorable performance overall in both the medium and low signal settings, and \textit{BJ-FG(true)} has the most favorable performance overall; these differences are especially noticeable in the low signal setting.

It is also clear that \textit{IPCW}$_2$ generally exhibits less variability than \textit{IPCW}$_1$ at time points $t_1$ and $t_2$, whereas for time $t_3$, the two measures perform similarly. This last result is expected since $\frac{\Delta(t^*)}{\hat G(T(t^*)|W)} \rightarrow \frac{\Delta}{\hat G(T|W)}$ as $t^* \rightarrow \infty$. 

The results further show that the choice of method for estimating the conditional expectation has little overall impact on performance; the main differences stem from the type of loss (IPCW versus DR versus BJ) and whether or not a composite loss function is used. The results also clearly highlight the value of getting the data generating model exactly right, and otherwise demonstrate a desirable degree of robustness across methods, particularly for doubly robust loss. When only comparing \textit{IPCW}$_{1}$ and the doubly robust methods, the efficiency gain is evident, particularly at time point $t_3$ using a single time point loss. The performance for IPCW$_2$ is often reasonably close to that for the Buckley-James and doubly robust loss functions, especially for a composite loss function; the 
comparative degree of simplicity involved in implementing this method has much to recommend it.

\begin{table}[ht]
\centering
\caption{{Numerical summaries for trees built using a single time point loss function when $n=250$. \textit{IPCW}$_{1}$ and \textit{IPCW}$_2$ are time-independent IPCW and time-dependent IPCW, respectively; \textit{BJ-RF(npar)}, \textit{BJ-RF+lgtc}, \textit{BJ-lgtc+godds} and \textit{BJ-FG(true)} are respectively built using Buckley-James loss functions with augmentation term estimated via nonparametric random forests, 
random parametric forests using ensembles of parametric models described in \cite{cheng2009modeling}, the parametric model of \cite{shi2013constrained},
and under the correct (i.e., consistently estimated) simulation model; and, 
\textit{DR-RF(npar)}, \textit{DR-RF+lgtc}, \textit{DR-lgtc+godds} and \textit{DR-FG(true)} are built using the doubly robust loss function, 
respectively using the same methods for calculating the augmentation term as in  \textit{BJ-RF(npar)}, \textit{BJ-RF+lgtc}, \textit{BJ-lgtc+godds} and \textit{BJ-FG(true)}.}}

\resizebox{\columnwidth}{!}{%
\begin{tabular}{rrrrrrrrrrr}
  \hline
	 & & \multicolumn{3}{c}{High Sig} & \multicolumn{3}{c}{Med Sig} & \multicolumn{3}{c}{Low Sig} \\ 
& & $|$fitted-3$|$ & NSP & PCSP & $|$fitted-3$|$ & NSP & PCSP & $|$fitted-3$|$ & NSP & PCSP \\ 
  \hline
\textit{IPCW}$_1$ & $t_1$ & 0.432 & 0.196 & 0.724 & 1.668 & 0.118 & 0.112 & 1.890 & 0.072 & 0.020 \\ 
 & $t_2$ & 0.204 & 0.138 & 0.890 & 0.838 & 0.166 & 0.494 & 1.644 & 0.124 & 0.118 \\ 
 & $t_3$ & 0.294 & 0.188 & 0.816 & 0.780 & 0.250 & 0.556 & 1.332 & 0.198 & 0.246 \\ \hline
 \textit{IPCW}$_2$ & $t_1$ & 0.350 & 0.230 & 0.800 & 1.428 & 0.208 & 0.220 & 1.834 & 0.114 & 0.040 \\ 
 & $t_2$ & 0.130 & 0.072 & 0.922 & 0.546 & 0.168 & 0.674 & 1.500 & 0.150 & 0.182 \\ 
 & $t_3$ & 0.354 & 0.242 & 0.800 & 0.726 & 0.216 & 0.576 & 1.318 & 0.134 & 0.238 \\ \hline
 \textit{BJ-RF(npar)} & $t_1$ & 0.392 & 0.280 & 0.814 & 1.298 & 0.286 & 0.284 & 1.840 & 0.160 & 0.042 \\ 
 & $t_2$ & 0.236 & 0.162 & 0.876 & 0.424 & 0.238 & 0.770 & 1.354 & 0.258 & 0.252 \\ 
 & $t_3$ & 0.320 & 0.258 & 0.858 & 0.336 & 0.218 & 0.818 & 0.980 & 0.304 & 0.438 \\ \hline
\textit{BJ-RF+lgtc} & $t_1$ & 0.326 & 0.234 & 0.822 & 1.302 & 0.304 & 0.290 & 1.840 & 0.140 & 0.044 \\ 
& $t_2$ & 0.256 & 0.182 & 0.864 & 0.404 & 0.228 & 0.774 & 1.344 & 0.232 & 0.256 \\ 
& $t_3$ & 0.302 & 0.236 & 0.858 & 0.234 & 0.144 & 0.854 & 0.930 & 0.256 & 0.448 \\ \hline
\textit{BJ-FG(true)} & $t_1$ & 0.336 & 0.260 & 0.826 & 0.950 & 0.284 & 0.482 & 1.694 & 0.152 & 0.106 \\ 
& $t_2$ & 0.190 & 0.140 & 0.902 & 0.204 & 0.134 & 0.882 & 0.682 & 0.246 & 0.630 \\ 
& $t_3$ & 0.142 & 0.116 & 0.932 & 0.150 & 0.098 & 0.908 & 0.202 & 0.138 & 0.886 \\ \hline
\textit{BJ-lgtc+godds} & $t_1$ & 0.392 & 0.282 & 0.792 & 1.182 & 0.320 & 0.354 & 1.766 & 0.146 & 0.070 \\ 
& $t_2$ & 0.342 & 0.244 & 0.850 & 0.368 & 0.236 & 0.818 & 1.024 & 0.284 & 0.404 \\ 
& $t_3$ & 0.616 & 0.336 & 0.728 & 0.426 & 0.262 & 0.816 & 0.702 & 0.386 & 0.602 \\ \hline
\textit{DR-RF(npar)} & $t_1$ & 0.342 & 0.248 & 0.814 & 1.378 & 0.246 & 0.246 & 1.838 & 0.138 & 0.044 \\ 
& $t_2$ & 0.170 & 0.128 & 0.898 & 0.508 & 0.242 & 0.696 & 1.506 & 0.222 & 0.182 \\ 
& $t_3$ & 0.288 & 0.188 & 0.860 & 0.480 & 0.128 & 0.688 & 1.382 & 0.122 & 0.236 \\ \hline
\textit{DR-RF+lgtc} & $t_1$ & 0.338 & 0.236 & 0.816 & 1.392 & 0.288 & 0.250 & 1.826 & 0.132 & 0.044 \\ 
& $t_2$ & 0.166 & 0.122 & 0.916 & 0.524 & 0.248 & 0.692 & 1.518 & 0.210 & 0.168 \\ 
& $t_3$ & 0.270 & 0.190 & 0.864 & 0.544 & 0.138 & 0.660 & 1.426 & 0.144 & 0.222 \\ \hline
\textit{DR-FG(true)} & $t_1$ & 0.366 & 0.266 & 0.818 & 1.362 & 0.268 & 0.274 & 1.830 & 0.134 & 0.046 \\ 
& $t_2$ & 0.168 & 0.126 & 0.910 & 0.482 & 0.208 & 0.734 & 1.416 & 0.150 & 0.212 \\ 
& $t_3$ & 0.244 & 0.164 & 0.866 & 0.588 & 0.234 & 0.686 & 1.318 & 0.148 & 0.284 \\ \hline
\textit{DR-lgtc+godds}& $t_1$ & 0.366 & 0.274 & 0.802 & 1.366 & 0.250 & 0.264 & 1.842 & 0.112 & 0.036 \\ 
& $t_2$ & 0.126 & 0.100 & 0.920 & 0.510 & 0.236 & 0.722 & 1.518 & 0.222 & 0.180 \\ 
& $t_3$ & 0.236 & 0.160 & 0.854 & 0.578 & 0.174 & 0.648 & 1.404 & 0.116 & 0.236 \\   \hline
\end{tabular}
}
\label{app-tab-1}
\end{table}

\begin{table}[ht]
\centering
\caption{{Numerical summaries for trees built using a single time point loss function when $n=500$. \textit{IPCW}$_{1}$ and \textit{IPCW}$_2$ are time-independent IPCW and time-dependent IPCW, respectively; \textit{BJ-RF(npar)}, \textit{BJ-RF+lgtc}, \textit{BJ-lgtc+godds} and \textit{BJ-FG(true)} are respectively built using Buckley-James loss functions with augmentation term estimated via nonparametric random forests, 
random parametric forests using ensembles of parametric models described in \cite{cheng2009modeling}, the parametric model of \cite{shi2013constrained},
and under the correct (i.e., consistently estimated) simulation model; and, 
\textit{DR-RF(npar)}, \textit{DR-RF+lgtc}, \textit{DR-lgtc+godds} and \textit{DR-FG(true)} are built using the doubly robust loss function, 
respectively using the same methods for calculating the augmentation term as in  \textit{BJ-RF(npar)}, \textit{BJ-RF+lgtc}, \textit{BJ-lgtc+godds} and \textit{BJ-FG(true)}.}}

\resizebox{\columnwidth}{!}{%
\begin{tabular}{rrrrrrrrrrr}
  \hline
	 & & \multicolumn{3}{c}{High Sig} & \multicolumn{3}{c}{Med Sig} & \multicolumn{3}{c}{Low Sig} \\ 
& & $|$fitted-3$|$ & NSP & PCSP & $|$fitted-3$|$ & NSP & PCSP & $|$fitted-3$|$ & NSP & PCSP \\ 
  \hline  
\textit{IPCW}$_1$ & $t_1$ & 0.154 & 0.110 & 0.892 & 1.058 & 0.184 & 0.422 & 1.842 & 0.078 & 0.064 \\ 
 & $t_2$ & 0.124 & 0.066 & 0.932 & 0.222 & 0.138 & 0.850 & 1.020 & 0.122 & 0.422 \\ 
 & $t_3$ & 0.134 & 0.092 & 0.916 & 0.178 & 0.106 & 0.892 & 0.672 & 0.134 & 0.594 \\ \hline
 \textit{IPCW}$_2$ & $t_1$ & 0.198 & 0.130 & 0.886 & 0.480 & 0.220 & 0.726 & 1.588 & 0.172 & 0.166 \\ 
 & $t_2$ & 0.118 & 0.076 & 0.928 & 0.120 & 0.090 & 0.904 & 0.666 & 0.170 & 0.614 \\ 
 & $t_3$ & 0.116 & 0.086 & 0.916 & 0.162 & 0.094 & 0.898 & 0.632 & 0.176 & 0.626 \\ \hline
 \textit{BJ-RF+npar} & $t_1$ & 0.192 & 0.128 & 0.906 & 0.394 & 0.208 & 0.774 & 1.506 & 0.192 & 0.214 \\ 
& $t_2$ & 0.120 & 0.070 & 0.924 & 0.178 & 0.126 & 0.894 & 0.366 & 0.162 & 0.784 \\ 
 & $t_3$ & 0.136 & 0.086 & 0.918 & 0.182 & 0.114 & 0.894 & 0.190 & 0.136 & 0.868 \\ \hline
 \textit{BJ-RF+lgtc} & $t_1$& 0.192 & 0.136 & 0.904 & 0.428 & 0.240 & 0.762 & 1.554 & 0.198 & 0.196 \\ 
 & $t_2$ & 0.086 & 0.052 & 0.934 & 0.170 & 0.120 & 0.898 & 0.362 & 0.162 & 0.792 \\ 
& $t_3$ & 0.142 & 0.090 & 0.918 & 0.174 & 0.108 & 0.900 & 0.200 & 0.140 & 0.870 \\ \hline
 \textit{BJ-FG(true)} & $t_1$  & 0.184 & 0.136 & 0.912 & 0.250 & 0.166 & 0.852 & 1.190 & 0.244 & 0.370 \\ 
 & $t_2$ & 0.110 & 0.072 & 0.950 & 0.186 & 0.134 & 0.880 & 0.122 & 0.086 & 0.906 \\ 
 & $t_3$ & 0.082 & 0.066 & 0.956 & 0.104 & 0.074 & 0.942 & 0.098 & 0.058 & 0.936 \\ \hline
\textit{BJ-lgtc+godds} & $t_1$ & 0.194 & 0.140 & 0.904 & 0.326 & 0.222 & 0.804 & 1.314 & 0.214 & 0.300 \\ 
& $t_2$ & 0.196 & 0.126 & 0.902 & 0.226 & 0.146 & 0.862 & 0.270 & 0.152 & 0.828 \\ 
& $t_3$ & 0.526 & 0.190 & 0.704 & 0.350 & 0.192 & 0.818 & 0.218 & 0.146 & 0.852 \\ \hline
\textit{DR-RF(npar)} & $t_1$ & 0.184 & 0.136 & 0.910 & 0.456 & 0.204 & 0.742 & 1.546 & 0.140 & 0.186 \\ 
& $t_2$ & 0.118 & 0.066 & 0.934 & 0.134 & 0.094 & 0.908 & 0.626 & 0.148 & 0.654 \\ 
& $t_3$ & 0.102 & 0.068 & 0.938 & 0.088 & 0.056 & 0.944 & 0.498 & 0.140 & 0.692 \\ \hline
\textit{DR-RF+lgtc} & $t_1$  & 0.158 & 0.116 & 0.918 & 0.482 & 0.202 & 0.722 & 1.568 & 0.146 & 0.180 \\ 
& $t_2$ & 0.108 & 0.052 & 0.930 & 0.118 & 0.088 & 0.916 & 0.644 & 0.158 & 0.652 \\ 
 & $t_3$ & 0.076 & 0.054 & 0.952 & 0.110 & 0.070 & 0.936 & 0.524 & 0.140 & 0.678 \\ \hline
\textit{DR-FG(true)} & $t_1$ & 0.190 & 0.140 & 0.906 & 0.444 & 0.188 & 0.756 & 1.576 & 0.172 & 0.186 \\ 
 & $t_2$ & 0.084 & 0.050 & 0.956 & 0.124 & 0.080 & 0.916 & 0.552 & 0.134 & 0.676 \\ 
& $t_3$ & 0.096 & 0.066 & 0.946 & 0.078 & 0.054 & 0.950 & 0.522 & 0.128 & 0.686 \\ \hline
\textit{DR-lgtc+godds} & $t_1$ & 0.206 & 0.140 & 0.904 & 0.440 & 0.186 & 0.744 & 1.552 & 0.192 & 0.170 \\ 
 & $t_2$ & 0.098 & 0.050 & 0.946 & 0.138 & 0.094 & 0.908 & 0.594 & 0.148 & 0.668 \\ 
 & $t_3$ & 0.084 & 0.054 & 0.948 & 0.118 & 0.076 & 0.928 & 0.496 & 0.112 & 0.682 \\   \hline
 \end{tabular}
}
\label{app-tab-2}
\end{table}

\begin{table}[ht]
\centering
\caption{{Numerical summaries for trees built using a single time point loss function when $n=1000$. \textit{IPCW}$_{1}$ and \textit{IPCW}$_2$ are time-independent IPCW and time-dependent IPCW, respectively; \textit{BJ-RF(npar)}, \textit{BJ-RF+lgtc}, \textit{BJ-lgtc+godds} and \textit{BJ-FG(true)} are respectively built using Buckley-James loss functions with augmentation term estimated via nonparametric random forests, 
random parametric forests using ensembles of parametric models described in \cite{cheng2009modeling}, the parametric model of \cite{shi2013constrained},
and under the correct (i.e., consistently estimated) simulation model; and, 
\textit{DR-RF(npar)}, \textit{DR-RF+lgtc}, \textit{DR-lgtc+godds} and \textit{DR-FG(true)} are built using the doubly robust loss function, 
respectively using the same methods for calculating the augmentation term as in  \textit{BJ-RF(npar)}, \textit{BJ-RF+lgtc}, \textit{BJ-lgtc+godds} and \textit{BJ-FG(true)}.}}

\resizebox{\columnwidth}{!}{%
\begin{tabular}{rrrrrrrrrrr}
  \hline
	 & & \multicolumn{3}{c}{High Sig} & \multicolumn{3}{c}{Med Sig} & \multicolumn{3}{c}{Low Sig} \\ 
& & $|$fitted-3$|$ & NSP & PCSP & $|$fitted-3$|$ & NSP & PCSP & $|$fitted-3$|$ & NSP & PCSP \\ 
  \hline 
 \textit{IPCW}$_1$ & $t_1$ & 0.106 & 0.076 & 0.930 & 0.132 & 0.072 & 0.908 & 1.328 & 0.134 & 0.284 \\ 
& $t_2$ & 0.066 & 0.042 & 0.962 & 0.134 & 0.114 & 0.904 & 0.208 & 0.116 & 0.864 \\ 
& $t_3$ & 0.092 & 0.054 & 0.936 & 0.070 & 0.052 & 0.952 & 0.124 & 0.088 & 0.916 \\ \hline
\textit{IPCW}$_2$ & $t_1$ & 0.088 & 0.072 & 0.946 & 0.092 & 0.070 & 0.934 & 0.646 & 0.170 & 0.650 \\ 
& $t_2$ & 0.064 & 0.042 & 0.956 & 0.116 & 0.090 & 0.924 & 0.122 & 0.076 & 0.916 \\ 
& $t_3$ & 0.100 & 0.056 & 0.932 & 0.082 & 0.056 & 0.944 & 0.130 & 0.094 & 0.922 \\ \hline
\textit{BJ-RF+npar} & $t_1$ & 0.106 & 0.082 & 0.924 & 0.114 & 0.084 & 0.932 & 0.488 & 0.156 & 0.722 \\ 
& $t_2$ & 0.088 & 0.056 & 0.950 & 0.168 & 0.118 & 0.890 & 0.108 & 0.070 & 0.924 \\ 
& $t_3$ & 0.114 & 0.060 & 0.932 & 0.180 & 0.110 & 0.894 & 0.142 & 0.106 & 0.906 \\ \hline
\textit{BJ-RF+lgtc} & $t_1$ & 0.126 & 0.102 & 0.922 & 0.104 & 0.078 & 0.938 & 0.458 & 0.152 & 0.738 \\ 
& $t_2$ & 0.064 & 0.048 & 0.958 & 0.136 & 0.094 & 0.904 & 0.086 & 0.056 & 0.934 \\ 
& $t_3$ & 0.124 & 0.068 & 0.924 & 0.160 & 0.104 & 0.898 & 0.138 & 0.094 & 0.904 \\ \hline
\textit{BJ-FG(true)} & $t_1$ & 0.108 & 0.088 & 0.932 & 0.088 & 0.056 & 0.946 & 0.218 & 0.100 & 0.866 \\ 
&$t_2$ & 0.084 & 0.070 & 0.956 & 0.058 & 0.042 & 0.954 & 0.076 & 0.050 & 0.944 \\ 
&$t_3$ & 0.036 & 0.024 & 0.970 & 0.104 & 0.074 & 0.942 & 0.060 & 0.040 & 0.958 \\ \hline
\textit{BJ-lgtc+godds} & $t_1$ & 0.112 & 0.084 & 0.922 & 0.110 & 0.082 & 0.934 & 0.302 & 0.120 & 0.820 \\ 
& $t_2$ & 0.164 & 0.094 & 0.922 & 0.162 & 0.090 & 0.902 & 0.100 & 0.076 & 0.932 \\ 
& $t_3$ & 0.774 & 0.128 & 0.508 & 0.358 & 0.126 & 0.786 & 0.204 & 0.122 & 0.860 \\ \hline
\textit{DR-RF(npar)} & $t_1$ & 0.088 & 0.070 & 0.942 & 0.102 & 0.072 & 0.934 & 0.556 & 0.168 & 0.692 \\ 
& $t_2$ & 0.068 & 0.048 & 0.958 & 0.098 & 0.068 & 0.930 & 0.104 & 0.072 & 0.926 \\ 
 & $t_3$ & 0.048 & 0.038 & 0.978 & 0.048 & 0.030 & 0.964 & 0.098 & 0.076 & 0.936 \\ \hline
\textit{DR-RF+lgtc} & $t_1$ & 0.092 & 0.070 & 0.938 & 0.094 & 0.066 & 0.938 & 0.588 & 0.168 & 0.682 \\ 
& $t_2$ & 0.048 & 0.034 & 0.962 & 0.100 & 0.068 & 0.932 & 0.090 & 0.064 & 0.934 \\ 
& $t_3$ & 0.040 & 0.032 & 0.982 & 0.034 & 0.024 & 0.974 & 0.098 & 0.074 & 0.938 \\ \hline
\textit{DR-FG(true)} & $t_1$ & 0.080 & 0.066 & 0.940 & 0.068 & 0.048 & 0.954 & 0.532 & 0.164 & 0.706 \\ 
& $t_2$ & 0.082 & 0.062 & 0.960 & 0.092 & 0.066 & 0.938 & 0.072 & 0.058 & 0.950 \\ 
& $t_3$ & 0.050 & 0.042 & 0.974 & 0.056 & 0.036 & 0.968 & 0.088 & 0.074 & 0.948 \\ \hline
\textit{DR-lgtc+godds} & $t_1$ & 0.068 & 0.054 & 0.948 & 0.094 & 0.064 & 0.938 & 0.564 & 0.162 & 0.690 \\ 
& $t_2$ & 0.054 & 0.036 & 0.962 & 0.114 & 0.078 & 0.934 & 0.074 & 0.054 & 0.938 \\ 
& $t_3$ & 0.048 & 0.022 & 0.970 & 0.040 & 0.024 & 0.972 & 0.062 & 0.052 & 0.948 \\ \hline
\end{tabular}
}
\label{app-tab-3}
\end{table}

\begin{table}[ht]
\centering
\caption{{Numerical summaries for trees built using multiple time points when $n=250, 500$ and $1000$. \textit{IPCW}$_{1}$ and \textit{IPCW}$_2$ are time-independent IPCW and time-dependent IPCW, respectively; \textit{BJ-RF(npar)}, \textit{BJ-RF+lgtc}, \textit{BJ-lgtc+godds} and \textit{BJ-FG(true)} are respectively built using Buckley-James loss functions with augmentation term estimated via nonparametric random forests, 
random parametric forests using ensembles of parametric models described in \cite{cheng2009modeling}, the parametric model of \cite{shi2013constrained},
and under the correct (i.e., consistently estimated) simulation model; and, 
\textit{DR-RF(npar)}, \textit{DR-RF+lgtc}, \textit{DR-lgtc+godds} and \textit{DR-FG(true)} are built using the doubly robust loss function, 
respectively using the same methods for calculating the augmentation term as in  \textit{BJ-RF(npar)}, \textit{BJ-RF+lgtc}, \textit{BJ-lgtc+godds} and \textit{BJ-FG(true)}.}}

\resizebox{\columnwidth}{!}{%
\begin{tabular}{crrrrrrrrrr}
  \hline
& & \multicolumn{3}{c}{High Sig} & \multicolumn{3}{c}{Med Sig} & \multicolumn{3}{c}{Low Sig} \\ 
& & $|$fitted-3$|$ & NSP & PCSP & $|$fitted-3$|$ & NSP & PCSP & $|$fitted-3$|$ & NSP & PCSP \\ 
  \hline
 $n=250$ & \textit{IPCW}$_1$ & 0.168 & 0.116 & 0.914 & 0.506 & 0.132 & 0.670 & 1.286 & 0.102 & 0.272 \\ 
& \textit{IPCW}$_2$ & 0.156 & 0.120 & 0.916 & 0.314 & 0.142 & 0.798 & 1.128 & 0.146 & 0.358 \\ 
& \textit{BJ-RF(npar)} & 0.234 & 0.186 & 0.878 & 0.304 & 0.158 & 0.810 & 1.128 & 0.246 & 0.364 \\ 
& \textit{BJ-RF+lgtc} & 0.250 & 0.196 & 0.882 & 0.290 & 0.158 & 0.824 & 1.110 & 0.244 & 0.360 \\ 
& \textit{BJ-FG(true)} & 0.156 & 0.116 & 0.910 & 0.194 & 0.142 & 0.892 & 0.348 & 0.182 & 0.790 \\ 
& \textit{BJ-lgtc+godds} & 0.358 & 0.246 & 0.844 & 0.364 & 0.266 & 0.814 & 0.788 & 0.328 & 0.532 \\ 
& \textit{DR-RF(npar)} & 0.150 & 0.120 & 0.924 & 0.272 & 0.138 & 0.828 & 1.156 & 0.146 & 0.334 \\ 
& \textit{DR-RF+lgtc} & 0.152 & 0.126 & 0.912 & 0.264 & 0.120 & 0.834 & 1.134 & 0.178 & 0.342 \\ 
& \textit{DR-FG(true)} & 0.132 & 0.090 & 0.932 & 0.194 & 0.092 & 0.880 & 1.050 & 0.170 & 0.408 \\ 
& \textit{DR-lgtc+godds} & 0.120 & 0.092 & 0.926 & 0.286 & 0.144 & 0.846 & 1.044 & 0.146 & 0.378 \\ \hline
 $n=500$ & \textit{IPCW}$_1$ & 0.132 & 0.084 & 0.916 & 0.182 & 0.132 & 0.874 & 0.558 & 0.164 & 0.658 \\ 
 & \textit{IPCW}$_2$ & 0.124 & 0.082 & 0.932 & 0.138 & 0.100 & 0.906 & 0.282 & 0.136 & 0.830 \\ 
 & \textit{BJ-RF(npar)}  & 0.142 & 0.090 & 0.904 & 0.148 & 0.088 & 0.908 & 0.226 & 0.128 & 0.878 \\ 
 & \textit{BJ-RF+lgtc} & 0.130 & 0.082 & 0.908 & 0.156 & 0.094 & 0.908 & 0.210 & 0.124 & 0.880 \\ 
 & \textit{BJ-FG(true)}  & 0.092 & 0.068 & 0.940 & 0.102 & 0.064 & 0.934 & 0.118 & 0.082 & 0.918 \\ 
 & \textit{BJ-lgtc+godds} & 0.250 & 0.122 & 0.848 & 0.156 & 0.080 & 0.896 & 0.154 & 0.104 & 0.896 \\ 
 & \textit{DR-RF(npar)} & 0.066 & 0.052 & 0.958 & 0.092 & 0.056 & 0.940 & 0.216 & 0.118 & 0.856 \\ 
 & \textit{DR-RF+lgtc}  & 0.040 & 0.030 & 0.966 & 0.092 & 0.052 & 0.938 & 0.230 & 0.106 & 0.834 \\ 
 & \textit{DR-FG(true)}  & 0.058 & 0.044 & 0.966 & 0.092 & 0.058 & 0.942 & 0.176 & 0.076 & 0.874 \\ 
 & \textit{DR-lgtc+godds} & 0.040 & 0.030 & 0.970 & 0.082 & 0.056 & 0.950 & 0.214 & 0.102 & 0.852 \\ \hline 
  $n=1000$ & \textit{IPCW}$_1$ & 0.080 & 0.036 & 0.940 & 0.122 & 0.096 & 0.928 & 0.088 & 0.074 & 0.936 \\ 
 & \textit{IPCW}$_2$ & 0.068 & 0.052 & 0.948 & 0.084 & 0.046 & 0.946 & 0.094 & 0.074 & 0.938 \\ 
  & \textit{BJ-RF(npar)}  & 0.110 & 0.062 & 0.918 & 0.112 & 0.074 & 0.918 & 0.096 & 0.068 & 0.934 \\ 
  & \textit{BJ-RF+lgtc} & 0.080 & 0.052 & 0.938 & 0.126 & 0.084 & 0.912 & 0.094 & 0.064 & 0.936 \\ 
  & \textit{BJ-FG(true)}  & 0.072 & 0.046 & 0.954 & 0.088 & 0.060 & 0.940 & 0.086 & 0.070 & 0.954 \\ 
 & \textit{BJ-lgtc+godds} & 0.270 & 0.078 & 0.818 & 0.140 & 0.076 & 0.904 & 0.100 & 0.066 & 0.928 \\ 
  & \textit{DR-RF(npar)} & 0.050 & 0.028 & 0.976 & 0.068 & 0.046 & 0.952 & 0.098 & 0.068 & 0.928 \\ 
 & \textit{DR-RF+lgtc}  & 0.038 & 0.022 & 0.978 & 0.056 & 0.044 & 0.956 & 0.104 & 0.078 & 0.930 \\ 
 & \textit{DR-FG(true)} & 0.030 & 0.022 & 0.978 & 0.056 & 0.034 & 0.966 & 0.058 & 0.042 & 0.958 \\ 
  & \textit{DR-lgtc+godds} & 0.028 & 0.016 & 0.984 & 0.060 & 0.044 & 0.952 & 0.082 & 0.056 & 0.944 \\ \hline
  \end{tabular}
}
\label{app-tab-4}
\end{table}

\clearpage

\subsection{Tree results in data analysis}
In this subsection, we show various trees built for the RTOG9410 study described in Section \ref{RTOG9410}. 
Numbers in each terminal node represent terminal node estimators (i.e., CIF estimates) generated
by the loss function used to build the corresponding tree.

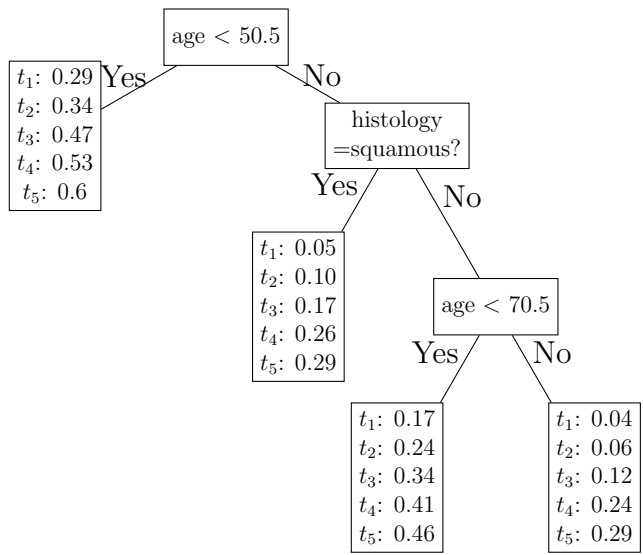
\begin{figure}[!ht]
\centering
\caption{Tree built using uncensored data for out-field failure}
\begin{tikzpicture}[scale=0.75]
\node [node,scale=0.75] (A) {age $<$ 50.5};
\path (A) ++(-150:\nodeDist) node [node,scale=0.75] (B) {$t_1$: 0.29\\$t_2$: 0.34\\$t_3$: 0.47\\$t_4$: 0.53\\$t_5$: 0.6};
\path (A) ++(-30:\nodeDist) node [node,scale=0.75] (C) {histology\\=squamous?};
\path (C) ++(-120:\nodeDist) node [node,scale=0.75] (D) {$t_1$: 0.05\\$t_2$: 0.10\\$t_3$: 0.17\\$t_4$: 0.26\\$t_5$: 0.29};
\path (C) ++(-60:\nodeDist) node [node,scale=0.75] (E) {age $<$ 70.5};
\path (E) ++(-120:\nodeDist) node [node,scale=0.75] (F) {$t_1$: 0.17\\$t_2$: 0.24\\$t_3$: 0.34\\$t_4$: 0.41\\$t_5$: 0.46};
\path (E) ++(-60:\nodeDist) node [node,scale=0.75] (G) {$t_1$: 0.04\\$t_2$: 0.06\\$t_3$: 0.12\\$t_4$: 0.24\\$t_5$: 0.29};

\draw (A) -- (B) node [left,pos=0.25] {Yes}(A);
\draw (A) -- (C) node [right,pos=0.25] {No}(A);
\draw (C) -- (D) node [left,pos=0.25] {Yes}(A);
\draw (C) -- (E) node [right,pos=0.25] {No}(A);
\draw (E) -- (F) node [left,pos=0.25] {Yes}(A);
\draw (E) -- (G) node [right,pos=0.25] {No}(A);
\end{tikzpicture}
\label{fig-of-u}
\end{figure}

\begin{figure}[!ht]
\caption{Tree built using 29\% artificially censored data for out-field failure for \textit{IPCW}$_1$}
\centering
\begin{tikzpicture}[scale=0.75]
\node [node,scale=0.75] (A) {age $<$ 50.5};
\path (A) ++(-150:\nodeDist) node [node,scale=0.75] (B) {$t_1$: 0.29 \\$t_2$ : 0.36 \\ $t_3$ : 0.50 \\ $t_4$:0.55 \\ $t_5$ : 0.61};
\path (A) ++(-30:\nodeDist) node [node,scale=0.75] (C) {age $<$ 71.5};
\path (C) ++(-60:\nodeDist) node [node,scale=0.75] (D) {$t_1$: 0.04 \\ $t_2$ : 0.04 \\ $t_3$ : 0.10 \\ $t_4$ : 0.12 \\ $t_5$: 0.14};
\path (C) ++(-120:\nodeDist) node [node,scale=0.75] (E) {histology\\=squamous?};
\path (E) ++(-120:\nodeDist) node [node,scale=0.75] (F) {$t_1$: 0.05 \\ $t_2$ : 0.11 \\ $t_3$: 0.17 \\ $t_4$: 0.26 \\ $t_5$ : 0.30};
\path (E) ++(-60:\nodeDist) node [node,scale=0.75] (G) {$t_1$: 0.16 \\ $t_2$ : 0.22 \\ $t_3$ : 0.33 \\ $t_4$ : 0.40 \\ $t_5$ : 0.46};

\draw (A) -- (B) node [left,pos=0.25] {Yes}(A);
\draw (A) -- (C) node [right,pos=0.25] {No}(A);
\draw (C) -- (D) node [right,pos=0.25] {No}(A);
\draw (C) -- (E) node [left,pos=0.25] {Yes}(A);
\draw (E) -- (F) node [left,pos=0.25] {Yes}(A);
\draw (E) -- (G) node [right,pos=0.25] {No}(A);
\end{tikzpicture}
\label{fig-of-outcen-ipcw1}
\end{figure}

\begin{figure}[!ht]
\caption{Tree built using 29\% artificially censored data for out-field failure for \textit{IPCW}$_2$}
\centering
\begin{tikzpicture}[scale=0.75]
\node [node,scale=0.75] (A) {age $<$ 50.5};
\path (A) ++(-150:\nodeDist) node [node,scale=0.75] (B) {$t_1$: 0.29 \\$t_2$ : 0.35 \\ $t_3$ : 0.48 \\ $t_4$:0.54  \\ $t_5$ : 0.61};
\path (A) ++(-30:\nodeDist) node [node,scale=0.75] (C) {age $<$ 71.5};
\path (C) ++(-60:\nodeDist) node [node,scale=0.75] (D) {$t_1$: 0.04 \\ $t_2$ : 0.04 \\ $t_3$ : 0.10 \\ $t_4$ : 0.12 \\ $t_5$: 0.15};
\path (C) ++(-120:\nodeDist) node [node,scale=0.75] (E) {histology\\=squamous?};
\path (E) ++(-120:\nodeDist) node [node,scale=0.75] (F) {$t_1$: 0.05 \\ $t_2$ : 0.11 \\ $t_3$: 0.17 \\ $t_4$: 0.25 \\ $t_5$ : 0.30};
\path (E) ++(-60:\nodeDist) node [node,scale=0.75] (G) {$t_1$: 0.16 \\ $t_2$ : 0.23 \\ $t_3$ : 0.33 \\ $t_4$ : 0.40 \\ $t_5$ : 0.46};

\draw (A) -- (B) node [left,pos=0.25] {Yes}(A);
\draw (A) -- (C) node [right,pos=0.25] {No}(A);
\draw (C) -- (D) node [right,pos=0.25] {No}(A);
\draw (C) -- (E) node [left,pos=0.25] {Yes}(A);
\draw (E) -- (F) node [left,pos=0.25] {Yes}(A);
\draw (E) -- (G) node [right,pos=0.25] {No}(A);
\end{tikzpicture}
\label{fig-of-outcen-ipcw2}
\end{figure}

\begin{figure}[!ht]
\caption{Tree built using 29\% artificially censored data for out-field failure for \textit{BJ-RF(npar)}} 
\centering
\begin{tikzpicture}[scale=0.75]
\node [node,scale=0.75] (A) {age $<$ 50.5};
\path (A) ++(-150:\nodeDist) node [node,scale=0.75] (B) {$t_1$: 0.28 \\ $t_2$ : 0.35 \\ $t_3$ : 0.47 \\ $t_4$ : 0.53 \\ $t_5$ : 0.58};
\path (A) ++(-30:\nodeDist) node [node,scale=0.75] (C) {age $<$ 71.5};
\path (C) ++(-60:\nodeDist) node [node,scale=0.75] (D) {$t_1$: 0.04 \\ $t_2$ : 0.04 \\ $t_3$ : 0.11 \\ $t_4$ : 0.14 \\ $t_5$ : 0.17};
\path (C) ++(-120:\nodeDist) node [node,scale=0.75] (E) {histology\\=squamous?};
\path (E) ++(-120:\nodeDist) node [node,scale=0.75] (F) {$t_1$: 0.06 \\ $t_2$ : 0.11 \\ $t_3$ : 0.18 \\ $t_4$ : 0.26 \\ $t_5$ : 0.30};
\path (E) ++(-60:\nodeDist) node [node,scale=0.75] (G) {$t_1$:  0.16 \\ $t_2$ : 0.23 \\ $t_3$ : 0.33 \\ $t_4$ :  0.40 \\ $t_5$ : 0.46};

\draw (A) -- (B) node [left,pos=0.25] {Yes}(A);
\draw (A) -- (C) node [right,pos=0.25] {No}(A);
\draw (C) -- (D) node [right,pos=0.25] {No}(A);
\draw (C) -- (E) node [left,pos=0.25] {Yes}(A);
\draw (E) -- (F) node [left,pos=0.25] {Yes}(A);
\draw (E) -- (G) node [right,pos=0.25] {No}(A);
\end{tikzpicture}
\label{fig-of-outcen-bj-npar}
\end{figure}

\begin{figure}[!ht]
\caption{Tree built using 29\% artificially censored data for out-field failure for \textit{DR-RF(npar)}} 
\centering
\begin{tikzpicture}[scale=0.75]
\node [node,scale=0.75] (A) {age $<$ 50.5};
\path (A) ++(-150:\nodeDist) node [node,scale=0.75] (B) {$t_1$: 0.29 \\ $t_2$ : 0.36 \\ $t_3$ : 0.49 \\ $t_4$ : 0.54 \\ $t_5$ : 0.60};
\path (A) ++(-30:\nodeDist) node [node,scale=0.75] (C) {age $<$ 71.5};
\path (C) ++(-60:\nodeDist) node [node,scale=0.75] (D) {$t_1$:0.04 \\ $t_2$ : 0.04 \\ $t_3$ : 0.10 \\ $t_4$ : 0.12 \\ $t_5$ : 0.15};
\path (C) ++(-120:\nodeDist) node [node,scale=0.75] (E) {histology\\=squamous?};
\path (E) ++(-120:\nodeDist) node [node,scale=0.75] (F) {$t_1$: 0.05 \\ $t_2$ : 0.11 \\ $t_3$ : 0.17 \\ $t_4$ : 0.25 \\ $t_5$ : 0.30};
\path (E) ++(-60:\nodeDist) node [node,scale=0.75] (G) {$t_1$: 0.16 \\ $t_2$ : 0.23 \\ $t_3$ : 0.34 \\ $t_4$ : 0.41 \\ $t_5$ :0.47 };

\draw (A) -- (B) node [left,pos=0.25] {Yes}(A);
\draw (A) -- (C) node [right,pos=0.25] {No}(A);
\draw (C) -- (D) node [right,pos=0.25] {No}(A);
\draw (C) -- (E) node [left,pos=0.25] {Yes}(A);
\draw (E) -- (F) node [left,pos=0.25] {Yes}(A);
\draw (E) -- (G) node [right,pos=0.25] {No}(A);
\end{tikzpicture}
\label{fig-of-outcen-dr-npar}
\end{figure}

\begin{figure}[!ht]
\centering
\caption{Tree built using 29\% artificially censored data for out-field failure for  \textit{BJ-RF+lgtc}}
\begin{tikzpicture}[scale=0.75]
\node [node,scale=0.75] (A) {age $<$ 50.5};
\path (A) ++(-150:\nodeDist) node [node,scale=0.75] (B) {$t_1$ : 0.27 \\ $t_2$ : 0.33 \\ $t_3$ :0.45 \\ $t_4$ :0.50 \\ $t_5$ :0.54};
\path (A) ++(-30:\nodeDist) node [node,scale=0.75] (C) {age $<$ 70.5};
\path (C) ++(-60:\nodeDist) node [node,scale=0.75] (D) { $t_1$ : 0.03 \\ $t_2$ : 0.03 \\ $t_3$ : 0.10 \\ $t_4$ : 0.12 \\ $t_5$ : 0.15};
\path (C) ++(-120:\nodeDist) node [node,scale=0.75] (E) {histology\\=squamous?};
\path (E) ++(-120:\nodeDist) node [node,scale=0.75] (F) {$t_1$ : 0.05\\ $t_2$ : 0.10 \\ $t_3$ : 0.15 \\ $t_4$ : 0.22 \\ $t_5$ : 0.26};
\path (E) ++(-60:\nodeDist) node [node,scale=0.75] (G) {$t_1$ : 0.16 \\ $t_2$ : 0.22 \\ $t_3$ : 0.31 \\ $t_4$ : 0.38 \\ $t_5$ : 0.42};

\draw (A) -- (B) node [left,pos=0.25] {Yes}(A);
\draw (A) -- (C) node [right,pos=0.25] {No}(A);
\draw (C) -- (D) node [right,pos=0.25] {No}(A);
\draw (C) -- (E) node [left,pos=0.25] {Yes}(A);
\draw (E) -- (F) node [left,pos=0.25] {Yes}(A);
\draw (E) -- (G) node [right,pos=0.25] {No}(A);
\end{tikzpicture}
\label{fig-of-outcen-bj-lgtc}
\end{figure}

\begin{figure}[!ht]
\centering
\caption{Tree built using 29\% artificially censored data for out-field failure for  \textit{BJ-RF+lgtc}}
\begin{tikzpicture}[scale=0.75]
\node [node,scale=0.75] (A) {age $<$ 50.5};
\path (A) ++(-150:\nodeDist) node [node,scale=0.75] (B) {$t_1$: 0.30 \\ $t_2$: 0.36 \\ $t_3$: 0.50 \\ $t_4$: 0.55 \\ $t_5$: 0.62};
\path (A) ++(-30:\nodeDist) node [node,scale=0.75] (C) {age $<$ 70.5};
\path (C) ++(-60:\nodeDist) node [node,scale=0.75] (D) { $t_1$ : 0.03 \\ $t_2$ : 0.03 \\ $t_3$ : 0.11 \\ $t_4$ : 0.13 \\ $t_5$ : 0.17};
\path (C) ++(-120:\nodeDist) node [node,scale=0.75] (E) {histology\\=squamous?};
\path (E) ++(-120:\nodeDist) node [node,scale=0.75] (F) {$t_1$: 0.05 \\ $t_2$: 0.11 \\ $t_3$: 0.16 \\ $t_4$: 0.25\\ $t_5$: 0.29};
\path (E) ++(-60:\nodeDist) node [node,scale=0.75] (G) {$t_1$ : 0.17 \\ $t_2$: 0.24 \\ $t_3$: 0.35 \\ $t_4$ : 0.42 \\ $t_5$: 0.48};

\draw (A) -- (B) node [left,pos=0.25] {Yes}(A);
\draw (A) -- (C) node [right,pos=0.25] {No}(A);
\draw (C) -- (D) node [right,pos=0.25] {No}(A);
\draw (C) -- (E) node [left,pos=0.25] {Yes}(A);
\draw (E) -- (F) node [left,pos=0.25] {Yes}(A);
\draw (E) -- (G) node [right,pos=0.25] {No}(A);
\end{tikzpicture}
\label{fig-of-outcen-dr-lgtc}
\end{figure}


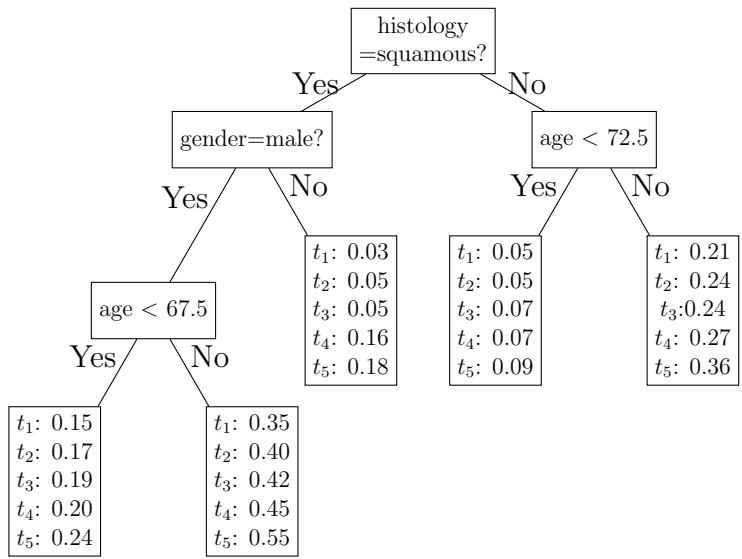
\begin{figure}[!ht]
\caption{Tree built using uncensored data for death without progression}
\centering
\begin{tikzpicture}[scale=0.75]
\node [node,scale=0.75] (A) {histology\\=squamous?};
\path (A) ++(-30:\nodeDist) node [node,scale=0.75] (B) {age $<$ 72.5};
\path (B) ++(-120:\nodeDist) node [node,scale=0.75] (C) {$t_1$: 0.05   \\ $t_2$: 0.05 \\ $t_3$: 0.07 \\ $t_4$: 0.07\\ $t_5$: 0.09};
\path (B) ++(-60:\nodeDist) node [node,scale=0.75] (D) {$t_1$: 0.21     \\$t_2$: 0.24 \\$t_3$:0.24  \\$t_4$: 0.27\\ $t_5$: 0.36};
\path (A) ++(-150:\nodeDist) node [node,scale=0.75] (E) {gender=male?};
\path (E) ++(-60:\nodeDist) node [node,scale=0.75] (F) {$t_1$:  0.03    \\$t_2$: 0.05 \\ $t_3$: 0.05\\$t_4$: 0.16\\$t_5$: 0.18};
\path (E) ++(-120:\nodeDist) node [node,scale=0.75] (G) {age $<$ 67.5};
\path (G) ++(-120:\nodeDist) node [node,scale=0.75] (H) {$t_1$: 0.15   \\$t_2$: 0.17  \\$t_3$: 0.19\\$t_4$: 0.20\\$t_5$: 0.24};
\path (G) ++(-60:\nodeDist) node [node,scale=0.75] (I) {$t_1$: 0.35  \\$t_2$: 0.40 \\$t_3$: 0.42\\$t_4$: 0.45\\$t_5$: 0.55};

\draw (A) -- (B) node [right,pos=0.25] {No}(A);
\draw (B) -- (C) node [left,pos=0.25] {Yes}(A);
\draw (B) -- (D) node [right,pos=0.25] {No}(A);
\draw (A) -- (E) node [left,pos=0.25] {Yes}(A);
\draw (E) -- (F) node [right,pos=0.25] {No}(A);
\draw (E) -- (G) node [left,pos=0.25] {Yes}(A);
\draw (G) -- (H) node [left,pos=0.25] {Yes}(A);
\draw (G) -- (I) node [right,pos=0.25] {No}(A);
\end{tikzpicture}
\label{fig-death-u}
\end{figure}

\begin{figure}[!ht]
\caption{Tree built using 29\% artificially censored data for death without progression from \textit{IPCW}$_2$}
\centering
\begin{tikzpicture}[scale=0.75]
\node [node,scale=0.75] (A) {histology\\=squamous?};
\path (A) ++(-30:\nodeDist) node [node,scale=0.75] (B) {age $<$ 70.5};
\path (B) ++(-120:\nodeDist) node [node,scale=0.75] (C) {$t_1$: 0.04 \\ $t_2$: 0.05\\$t_3$: 0.06\\ $t_4$: 0.07 \\ $t_5$: 0.08};
\path (B) ++(-60:\nodeDist) node [node,scale=0.75] (D) {$t_1$: 0.20 \\ $t_2$: 0.23\\ $t_3$: 0.24\\ $t_4$: 0.24\\ $t_5$: 0.31};
\path (A) ++(-150:\nodeDist) node [node,scale=0.75] (E) {gender=male?};
\path (E) ++(-60:\nodeDist) node [node,scale=0.75] (F) {$t_1$: 0.03 \\ $t_2$: 0.05 \\ $t_3$: 0.05\\ $t_4$: 0.16\\$t_5$: 0.16 };
\path (E) ++(-120:\nodeDist) node [node,scale=0.75] (G) {age $<$ 67.5};
\path (G) ++(-120:\nodeDist) node [node,scale=0.75] (H) {$t_1$: 0.16 \\ $t_2$: 0.17 \\ $t_3$: 0.19 \\ $t_4$: 0.20\\ $t_5$: 0.23};
\path (G) ++(-60:\nodeDist) node [node,scale=0.75] (I) {$t_1$: 0.36 \\ $t_2$: 0.43 \\ $t_3$: 0.46 \\ $t_4$: 0.48 \\ $t_5:$ 0.62};

\draw (A) -- (B) node [right,pos=0.25] {No}(A);
\draw (B) -- (C) node [left,pos=0.25] {Yes}(A);
\draw (B) -- (D) node [right,pos=0.25] {No}(A);
\draw (A) -- (E) node [left,pos=0.25] {Yes}(A);
\draw (E) -- (F) node [right,pos=0.25] {No}(A);
\draw (E) -- (G) node [left,pos=0.25] {Yes}(A);
\draw (G) -- (H) node [left,pos=0.25] {Yes}(A);
\draw (G) -- (I) node [right,pos=0.25] {No}(A);
\end{tikzpicture}
\label{fig-death-cen-ipcw2}
\end{figure}

\begin{figure}[!ht]
\caption{Tree built using 29\% artificially censored data for death without progression from \textit{DR-RF(npar)} }
\centering
\begin{tikzpicture}[scale=0.75]
\node [node,scale=0.75] (A) {histology\\=squamous?};
\path (A) ++(-30:\nodeDist) node [node,scale=0.75] (B) {age $<$ 72.5};
\path (B) ++(-120:\nodeDist) node [node,scale=0.75] (C) {$t_1$: 0.05 \\ $t_2$ : 0.05 \\ $t_3$: 0.07\\ $t_4$: 0.07 \\ $t_5$: 0.08};
\path (B) ++(-60:\nodeDist) node [node,scale=0.75] (D) {$t_1$: 0.21 \\ $t_2$: 0.25\\ $t_3$: 0.25\\$t_4$: 0.25\\$t_5$: 0.34};
\path (A) ++(-150:\nodeDist) node [node,scale=0.75] (E) {gender=male?};
\path (E) ++(-60:\nodeDist) node [node,scale=0.75] (F) {$t_1$: 0.03 \\ $t_2$: 0.05 \\ $t_3$: 0.05 \\ $t_4$: 0.15 \\ $t_5$: 0.16 };
\path (E) ++(-120:\nodeDist) node [node,scale=0.75] (G) {age $<$ 67.5};
\path (G) ++(-120:\nodeDist) node [node,scale=0.75] (H) {$t_1$: 0.15 \\ $t_2$: 0.16 \\ $t_3$: 0.19 \\ $t_4$: 0.20 \\ $t_5$ : 0.23};
\path (G) ++(-60:\nodeDist) node [node,scale=0.75] (I) {$t_1$: 0.35 \\ $t_2$: 0.41 \\ $t_3$: 0.44 \\ $t_4:$ 0.44 \\ $t_5$: 0.56};

\draw (A) -- (B) node [right,pos=0.25] {No}(A);
\draw (B) -- (C) node [left,pos=0.25] {Yes}(A);
\draw (B) -- (D) node [right,pos=0.25] {No}(A);
\draw (A) -- (E) node [left,pos=0.25] {Yes}(A);
\draw (E) -- (F) node [right,pos=0.25] {No}(A);
\draw (E) -- (G) node [left,pos=0.25] {Yes}(A);
\draw (G) -- (H) node [left,pos=0.25] {Yes}(A);
\draw (G) -- (I) node [right,pos=0.25] {No}(A);
\end{tikzpicture}
\label{fig-death-cen-dr-rf-npar}
\end{figure}

\begin{figure}[!ht]
\caption{Tree built using 29\% artificially censored data for death without progression from \textit{DR-RF+lgtc}}
\centering
\begin{tikzpicture}[scale=0.75]
\node [node,scale=0.75] (A) {histology\\=squamous?};
\path (A) ++(-30:\nodeDist) node [node,scale=0.75] (B) {age $<$ 70.5};
\path (B) ++(-120:\nodeDist) node [node,scale=0.75] (C) {$t_1$: 0.04 \\ $t_2$ : 0.05 \\ $t_3$: 0.06\\ $t_4$: 0.06\\ $t_5$: 0.07 };
\path (B) ++(-60:\nodeDist) node [node,scale=0.75] (D) {$t_1$: 0.19 \\ $t_2$: 0.22 \\ $t_3$: 0.22 \\ $t_4$: 0.23 \\ $t_5$: 0.30  };
\path (A) ++(-150:\nodeDist) node [node,scale=0.75] (E) {gender=male?};
\path (E) ++(-60:\nodeDist) node [node,scale=0.75] (F) {$t_1$: 0.03 \\ $t_2$: 0.05 \\ $t_3$: 0.05 \\ $t_4$: 0.15 \\ $t_5$: 0.15 };
\path (E) ++(-120:\nodeDist) node [node,scale=0.75] (G) {age $<$ 67.5};
\path (G) ++(-120:\nodeDist) node [node,scale=0.75] (H) {$t_1$: 0.15 \\ $t_2$: 0.16 \\ $t_3$: 0.19 \\ $t_4$: 0.20 \\ $t_5$: 0.23};
\path (G) ++(-60:\nodeDist) node [node,scale=0.75] (I) {$t_1$: 0.35 \\ $t_2$: 0.41 \\ $t_3$: 0.44 \\ $t_4$: 0.44 \\ $t_5$: 0.57};

\draw (A) -- (B) node [right,pos=0.25] {No}(A);
\draw (B) -- (C) node [left,pos=0.25] {Yes}(A);
\draw (B) -- (D) node [right,pos=0.25] {No}(A);
\draw (A) -- (E) node [left,pos=0.25] {Yes}(A);
\draw (E) -- (F) node [right,pos=0.25] {No}(A);
\draw (E) -- (G) node [left,pos=0.25] {Yes}(A);
\draw (G) -- (H) node [left,pos=0.25] {Yes}(A);
\draw (G) -- (I) node [right,pos=0.25] {No}(A);
\end{tikzpicture}
\label{fig-death-cen-dr-rf-lgtc}
\end{figure}

\begin{figure}[!ht]
\centering
\caption{Tree built using 29\% artificially censored data for death without progression from \textit{BJ-RF(npar)} }
\begin{tikzpicture}[scale=0.75]
\node [node,scale=0.75] (A) {Histology\\=squamous?};
\path (A) ++(-30:\nodeDist) node [node,scale=0.75] (B) {age $<$ 72.5};
\path (B) ++(-120:\nodeDist) node [node,scale=0.75] (C) {$t_1$: 0.05 \\ $t_2$ : 0.05 \\ $t_3$: 0.07 \\ $t_4$ : 0.08 \\ $t_5$: 0.09};
\path (B) ++(-60:\nodeDist) node [node,scale=0.75] (D) {$t_1$: 0.22 \\ $t_2$: 0.25 \\ $t_3$: 0.25 \\ $t_4$ : 0.25 \\ $t_5$:0.33};
\path (A) ++(-150:\nodeDist) node [node,scale=0.75] (E) {gender=male?};
\path (E) ++(-60:\nodeDist) node [node,scale=0.75] (F) {$t_1$: 0.04 \\ $t_2$: 0.05 \\ $t_3$: 0.06 \\ $t_4$: 0.15 \\ $t_5$: 0.15};
\path (E) ++(-120:\nodeDist) node [node,scale=0.75] (G) {age $<$ 67.5};
\path (G) ++(-60:\nodeDist) node [node,scale=0.75] (H) {$t_1$: 0.33 \\ $t_2$: 0.38\\ $t_3$: 0.41 \\$t_4$: 0.42\\ $t_5$: 0.51};
\path (G) ++(-120:\nodeDist) node [node,scale=0.75] (I) {kps $<$ 95};
\path (I) ++(-120:\nodeDist) node [node,scale=0.75] (J) {$t_1$: 0.18 \\ $t_2$: 0.20\\ $t_3$: 0.23\\ $t_4$: 0.24\\ $t_5$: 0.26};
\path (I) ++(-60:\nodeDist) node [node,scale=0.75] (K) {$t_1$: 0.04 \\ $t_2$: 0.05 \\ $t_3$: 0.05 \\$t_4$: 0.06\\ $t_5$: 0.10}; 

\draw (A) -- (B) node [right,pos=0.25] {No}(A);
\draw (B) -- (C) node [left,pos=0.25] {Yes}(A);
\draw (B) -- (D) node [right,pos=0.25] {No}(A);
\draw (A) -- (E) node [left,pos=0.25] {Yes}(A);
\draw (E) -- (F) node [right,pos=0.25] {No}(A);
\draw (E) -- (G) node [left,pos=0.25] {Yes}(A);
\draw (G) -- (H) node [right,pos=0.25] {No}(A);
\draw (G) -- (I) node [left,pos=0.25] {Yes}(A);
\draw (I) -- (J) node [left,pos=0.25] {Yes}(A);
\draw (I) -- (K) node [right,pos=0.25] {No}(A);
\end{tikzpicture}
\label{fig-death-cen-bj-rf-npar}
\end{figure}

\begin{figure}[!ht]
\centering
\caption{Tree built using 29\% artificially censored data for death without progression from \textit{BJ-RF+lgtc}}
\begin{tikzpicture}[scale=0.75]
\node [node,scale=0.75] (A) {age $<$ 67.5};
\path (A) ++(-30:\nodeDist) node [node,scale=0.75] (B) {histology\\=squamous?};
\path (B) ++(-60:\nodeDist) node [node,scale=0.75] (C) {$t_1$: 0.05 \\ $t_2$: 0.05 \\ $t_3$: 0.07 \\ $t_4$: 0.09 \\ $t_5$: 0.12 };
\path (B) ++(-120:\nodeDist) node [node,scale=0.75] (D) {$t_1$: 0.12 \\ $t_2$: 0.14 \\ $t_3$: 0.16 \\ $t_4$: 0.21 \\ $t_5$: 0.25 };
\path (A) ++(-150:\nodeDist) node [node,scale=0.75] (E) {gender=male?};
\path (E) ++(-60:\nodeDist) node [node,scale=0.75] (F) {$t_1$: 0.03 \\ $t_2$: 0.03 \\ $t_3$: 0.04 \\ $t_4$: 0.06 \\ $t_5$: 0.12};
\path (E) ++(-120:\nodeDist) node [node,scale=0.75] (G) {kps $<$ 85};
\path (G) ++(-120:\nodeDist) node [node,scale=0.75] (H) {$t_1$: 0.46 \\ $t_2$: 0.51 \\ $t_3$: 0.52 \\ $t_4$: 0.54 \\ $t_5$: 0.56};
\path (G) ++(-60:\nodeDist) node [node,scale=0.75] (I) {RX=1};
\path (I) ++(-120:\nodeDist) node [node,scale=0.75] (J) {$t_1$: 0.12 \\ $t_2$: 0.15 \\ $t_3$: 0.16 \\ $t_4$: 0.18 \\ $t_5$: 0.26};
\path (I) ++(-60:\nodeDist) node [node,scale=0.75] (K) {$t_1$: 0.32 \\ $t_2$: 0.35 \\ $t_3$: 0.43 \\ $t_4$: 0.45 \\ $t_5$: 0.56};

\draw (A) -- (B) node [right,pos=0.25] {No}(A);
\draw (B) -- (C) node [right,pos=0.25] {No}(A);
\draw (B) -- (D) node [left,pos=0.25] {Yes}(A);
\draw (A) -- (E) node [left,pos=0.25] {Yes}(A);
\draw (E) -- (F) node [right,pos=0.25] {No}(A);
\draw (E) -- (G) node [left,pos=0.25] {Yes}(A);
\draw (G) -- (H) node [left,pos=0.25] {Yes}(A);
\draw (G) -- (I) node [right,pos=0.25] {No}(A);
\draw (I) -- (J) node [left,pos=0.25] {Yes}(A);
\draw (I) -- (K) node [right,pos=0.25] {No}(A);
\end{tikzpicture}
\label{fig-death-cen-bj-rf-lgtc}
\end{figure}

\clearpage
\newpage

\singlespace
\bibliographystyle{apalike}
\bibliography{refsall}

\end{document}